\documentclass[twocolumn,twocolappendix]{aastex701}

\usepackage[dvipsnames,svgnames]{xcolor} 
\definecolor{mygreen}{rgb}{0.55,0.71,0.0}
\definecolor{amethyst}{rgb}{0.6, 0.4, 0.8}
\definecolor{alizarin}{rgb}{0.82, 0.1, 0.26}
\definecolor{green}{rgb}{0.55, 0.71, 0.0}
\definecolor{apricot}{rgb}{0.98, 0.81, 0.69}
\definecolor{auburn}{rgb}{0.43, 0.21, 0.1}
\definecolor{babyblueeyes}{rgb}{0.63, 0.79, 0.95}
\definecolor{bittersweet}{rgb}{1.0, 0.44, 0.37}
\definecolor{arsenic}{rgb}{0.23, 0.27, 0.29}

\usepackage[normalem]{ulem}
\newcommand{\cfsout}{\bgroup\markoverwith{\textcolor{red}{\rule[0.5ex]{2pt}{0.4pt}}}\ULon}

\usepackage{graphicx}	
\usepackage{amsmath} 
\usepackage{amssymb}
\usepackage{gensymb}

\usepackage{verbatim}
\usepackage{longtable}
\usepackage{booktabs}    
\usepackage{multirow}

\usepackage{array}
\newcolumntype{L}[1]{>{\raggedright\arraybackslash}p{#1}}

\input ArtNouvc.fd
\newcommand*\initfamily{\usefont{U}{ArtNouvc}{xl}{n}}
  {\initfamily}%
  {}

\newcommand{\gray}{$\gamma$-ray}

\newcommand{\fcat}{\texttt{1FLAT}}
\newcommand{\f}{\texttt{Firmamento}}
\newcommand{\fshort}{\raisebox{-0.5ex}{\includegraphics[height=1em]{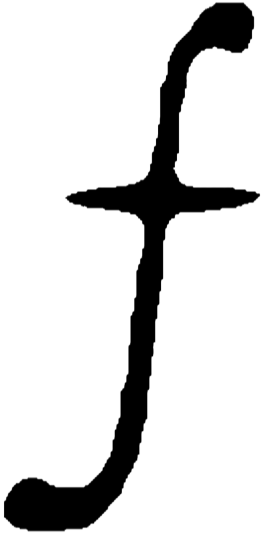}}}
\newcommand{\fermi}{\textit {Fermi}-LAT}
\newcommand{\lat}{LAT}

\usepackage{hyperref,url}

\shorttitle{The \fcat\ catalog}
\shortauthors{Giommi, P. et al.}

\accepted{October 14, 2025}
\submitjournal{ApJS}
\graphicspath{{./}{Figures/}}

\begin{document}

\title{\fcat : a \f -based catalog of AGN in \fermi\ high Galactic latitude \gray\ sources}

\author[orcid=0000-0002-2265-5003]{P. Giommi}
\affiliation{Center for Astrophysics and Space Science (CASS), New York University-Abu-Dhabi, P.O. Box 129188 Abu Dhabi, United Arab Emirates}
\affiliation{INAF, Brera Astronomical Observatory, via Brera, 28, I-20121 Milano, Italy}
\email[show]{giommipaolo@gmail.com}

\author[orcid=0000-0001-9104-3214]{M. Doro}
\affiliation{Department of Physics and Astronomy, University of Padova, via Marzolo 8, I-35131 Padova, Italy}
\affiliation{Istituto Nazionale Fisica Nucleare (INFN), Sezione di Padova, via Marzolo 8, I-35131 Padova, Italy}
\email[show]{michele.doro@unipd.it}  

\author[orcid=0009-0005-2763-0023]{M. Gouvêa}
\affiliation{Centro Brasileiro de Pesquisas Físicas, Rua Dr. Xavier Sigaud 150, 22290-180, Rio de Janeiro, Brazil}
\email[]{marcela.gouvearrb@gmail.com}

\author[orcid=0009-0009-3465-2035]{L. Fronte}
\affiliation{Department of Physics, University of Pisa, via Buonarroti 3, I-56127 Pisa, Italy
}
\email[]{laurafronte2004@gmail.com}

\author[orcid=0009-0002-7085-2244]{F. Metruccio}
\affiliation{Department of Physics and Astronomy, University of Padova, via Marzolo 8, I-35131 Padova, Italy}
\email[]{francesco.metruccio@studenti.unipd.it}

\author[orcid=0000-0002-1061-0510]{F. Arneodo}
\affiliation{New York University-Abu Dhabi, P.O. Box 129188 Abu Dhabi, United Arab Emirates}
\email{francesco.arneodo@nyu.edu}

\author[orcid=0000-0001-7909-588X]{U. Barres de Almeida}
\affiliation{Centro Brasileiro de Pesquisas Físicas, Rua Dr. Xavier Sigaud 150, 22290-180, Rio de Janeiro, Brazil}
\email{ulisses@cbpf.br}

\author[orcid=0000-0002-9744-071X]{S. Di Pippo}
\affiliation{New York University-Abu Dhabi, P.O. Box 129188 Abu Dhabi, United Arab Emirates}
\affiliation{SDA Bocconi, Via Sarfatti 25, I-20100, Milano, Italy}
\email{simonetta.dipippo@gmail.com}

\author{T. Kerscher}
\affiliation{Technical University of Munich, TUM School of Natural Sciences, Physics Department, 85747 Garching, Germany
}
\email[]{tobias.kerscher@tum.de}

\author[orcid=0009-0004-4596-7941]{A. Macciò}
\affiliation{Center for Astrophysics and Space Science (CASS), New York University-Abu-Dhabi, P.O. Box 129188 Abu Dhabi, United Arab Emirates}
\email[]{avm4@nyu.edu}

\author[orcid=0009-0005-7020-7976]{B. Mazzon}
\affiliation{Department of Environmental Sciences, Statistics and Computer Science, Ca' Foscari University of Venice, via Torino 155, I-30170 Venezia, Italy}
\email[]{mazzon.brando@gmail.com}

\author[orcid=0009-0004-0999-1098]{M. Morrone}
\affiliation{Department of Physics and Astronomy, University of Padova, via Marzolo 8, I-35131 Padova, Italy}
\email[]{martina.morrone@studenti.unipd.it}

\author[orcid=0000-0003-4502-9053]{E. Prandini}
\affiliation{Department of Physics and Astronomy, University of Padova, via Marzolo 8, I-35131 Padova, Italy}
\affiliation{Istituto Nazionale Fisica Nucleare (INFN), Sezione di Padova, via Marzolo 8, I-35131 Padova, Italy}
\email[]{elisa.prandini@unipd.it}

\author{A. Rodríguez}
\affiliation{New York University-Abu Dhabi, P.O. Box 129188 Abu Dhabi, United Arab Emirates}
\email[]{aer9873@nyu.edu}

\author[orcid=0000-0001-6708-6580]{A. Ruina}
\affiliation{Istituto Nazionale Fisica Nucleare (INFN), Sezione di Padova, via Marzolo 8, I-35131 Padova, Italy}
\email[]{arshia.ruina@pd.infn.it}

\author[orcid=0000-0003-2011-2731]{N. Sahakyan}
\affiliation{ICRANet-Armenia, Marshall Baghramian Avenue 24a, Yerevan 0019, Armenia}
\email[]{narsahakyan@gmail.com}

\author[orcid=0000-0002-6825-714X]{L.~Silveri}
\affiliation{New York University-Abu Dhabi, P.O. Box 129188 Abu Dhabi, United Arab Emirates}
\email[]{leandro.silveri@nyu.edu }

\author[orcid=0000-0003-1689-6254]{D. Tripathi}
\affiliation{New York University-Abu Dhabi, P.O. Box 129188 Abu Dhabi, United Arab Emirates}
\email[]{dt2202@nyu.edu}

\begin{abstract}
We present a systematic reassessment of 5,062 high–Galactic latitude gamma-ray sources from the \fermi\ 4FGL-DR4  catalog using \f\,, a web-based platform for multi-frequency source discovery and analysis. 
Our goal is to provide an independent evaluation of LAT \gray\, source associations using alternative spectral and spatial methods that integrate both recent and legacy survey data.
The evaluation is further refined through human supervision of SEDs, 
source morphology, flux variability, and template-based comparisons.
\f\ confirms the 4FGL-DR4  and 4LAC-DR3 counterparts or unassociated sources in 4,493 cases (88.8\%), demonstrating the robustness of both approaches. Beyond this general agreement, we identify 421 new blazar counterparts among previously unassociated sources, thereby reducing the fraction of unidentified extragalactic \fermi\ sources from 25\% to 17\%. In addition, in 64  cases we find alternative blazar associations, while in 49 instances,  we do not confirm the 4FGL-DR4 association. For all confirmed blazar counterparts we provide homogeneous estimates of synchrotron peak frequency and peak flux using machine-learning and template-based methods. The results agree with 4LAC-DR3 values in most cases, though significant discrepancies appear for a few dozen sources, often due to improved X-ray coverage.
The primary outcome of this work is the 1st \f\ LAT AGN Table (\fcat ), made publicly available through the \f\ platform (\url{https://firmamento.nyuad.nyu.edu}), where all related multi-wavelength data and images are available. The project involved extensive manual validation and benefited from the active participation of graduate and undergraduate students, highlighting the platform’s value for both research and education.
\end{abstract}



\vspace*{1cm}
\section{Introduction}
\label{sec:introduction}

A small fraction of active galactic nuclei (AGN) launch powerful relativistic jets that emit radiation across the entire electromagnetic spectrum, making them prominent sources in $\gamma$-ray surveys~\citep{beckmann2012agn,dermer2009high}. When these jets are aligned with the observer’s line of sight, the resulting relativistic Doppler boosting amplifies the jet’s luminosity, often outshining other emission components of the AGN and leading to their classification as blazars ~\citep{Padovani:2017}. Blazars exhibit a characteristic multi-wavelength emission, with a radio-to-\gray\, Spectral Energy Distribution (SED) that typically shows two broad components \citep{2010Abdo}. The low-energy component, spanning from radio to optical frequencies, and sometimes extending to the X-ray band, is attributed to synchrotron radiation from relativistic electrons spiraling in magnetic fields. The high-energy emission (X-rays to gamma-rays) arises from inverse Compton scattering of lower-energy photons or from other non-thermal processes such as hadronic interactions.

Blazars are classified according to the shape of their SEDs, characterized by the location of the synchrotron peak ($\nu_{\text{peak}}$), which sheds light on the power budget of the AGN~\citep{padovani1995,2010Abdo}. In this work we classify blazars as Low-Synchrotron Peaked (LSP) if $\nu_{\text{peak}} < 10^{13.5}$ Hz, Intermediate-Synchrotron Peaked (ISP) if $10^{13.5} \leq \nu_{\text{peak}} < 10^{15}$ Hz and High-Synchrotron Peaked (HSP) if $\nu_{\text{peak}} \geq 10^{15}$ Hz. 

Large-scale $\gamma$-ray surveys conducted by telescopes such as \fermi\ (hereafter \lat) have resulted in extensive source catalogs, the most recent being the fourth catalog, 4th data release (4FGL-DR4, hereafter 4FGL)~\citep{4FGL-DR3,4FGL-DR4}, which builds up from the first release 4FGL-DR1~\citep{4FGL-DR1}. The catalog provides extensive information on source detection, including coordinates, significance, spectral fits, etc, and additionally provides proposed counterparts. Such associations are based on unsupervised Bayesian and Likelihood Ratio (LR) statistics algorithms. The Bayesian method is based solely on spatial coincidence between the gamma-ray sources and their potential counterparts, the LR method using their $\log{N}–\log{S}$. The probability of association is also estimated using a prior based on the number of emitters in the error circle, and the association is retained if the probability \texttt{ASSOC\_PROB\_BAY} is $\geq 0.8$~\citep{4FGL-DR3}. A large fraction of 4FGL sources were separately further investigated specifically in search for blazar counterparts, making up the dedicated 4LAC-DR3 catalog~\citep{4LAC-DR3} (hereafter 4LAC). The 4FGL catalog contains 7,194 sources, 5,062 of which are located at Galactic latitude $|b|\geq10^\circ$, while 4LAC reports 3,407  blazars in this region (67\% of the 4FGL sources).

Although these catalogs have significantly advanced our understanding of the $\gamma$-ray sky, a considerable fraction of sources, approximately 25\% in the case of 4FGL, remain unidentified due to the lack of clear counterparts at other wavelengths. Among those, \citet{4FGL-DR3} report that it is very likely that those at high Galactic latitudes are likely unassociated blazars.
Identifying the nature of these sources is crucial for completing the census of $\gamma$-ray emitters and potentially discovering new classes of astrophysical objects.

To address this challenge, we present a systematic reassessment of the AGN counterparts of high-Galactic latitude $\gamma$-ray sources in the 4FGL, utilizing \f~\citep{firmamento} (hereafter \fshort), a novel web-based platform specifically developed for the search and identification of multi-frequency counterparts of X-ray and $\gamma$-ray sources. \fshort\ employs advanced data-handling capabilities, including machine learning models and specialized data science tools~\citep{VOU-Blazars,blast,wpeak}. Our search is primarily spectromorphological. We obtain multi-wavelength data through a survey search (see \autoref{app:survey}), compare the spatial morphology and the SED relations to match candidates~\citep{firmamento}.

With respect to 4FGL, we directly access multi-wavelength survey data, rather than source catalogs, including  a number of recent surveys that were not available when 4FGL and 4LAC were prepared. Another important difference is that we include source-by-source human validation on the SED as well as the multi-wavelength sky maps to supervise the association. Because of this, an accurate false-positive rate cannot be firmly statistically evaluated, although we discuss the robustness of our results throughout the work.  Furthermore, in line with \cite{4LAC-DR3} we compute the synchrotron peak and flux at this frequency and compare these values with those of 4LAC. This massive manual work  benefited from the active participation of graduate and undergraduate students through \fshort’s user-friendly interface and commitment to educational engagement.

\bigskip
The structure of this paper is as follows. \autoref{sec:methodology} provides a concise summary of the features of the \fshort\ platform (presented extensively in \citep{firmamento}) used for this work. \autoref{sec:results} presents the main numerical results of our analysis, discussing both the agreement and discrepancies with previous \lat\ catalogs and introduce the \fcat\, catalog. In \autoref{sec:analysis} we study more in depth the properties of our classification, investigating the possible reasons for discrepancies with 4FGL. In \autoref{sec:citizen} we discuss the educational engagement utilized in this project. Finally, in \autoref{sec:conclusion} we discuss and summarize our findings.

\section{Methodology}
\label{sec:methodology}

\subsection{The Firmamento Platform} 
\label{sec:firmamento}
\fshort\ is a novel, web-based data analysis platform designed for the discovery and study of multi-frequency and multi-messenger astrophysical sources \citep{firmamento}. It provides a comprehensive suite of tools for exploring sources across the electromagnetic spectrum, integrating extensive multi-band catalogs with advanced data-handling capabilities, including machine learning, through an accessible visual interface \citep{Giommi2025}, and it is conceived for educational engagement. \fshort\ is undergoing constant development to add features, extend investigation to different astrophysical classes of objects, improve performance and appearance. \fshort\ is developed in the framework of the Open Universe initiative --- an effort under the auspices of the United Nations Committee on the Peaceful Uses of Outer Space (COPUOS) and implemented by United Nations Office for Outer Space Affairs (UNOOSA)  \citep{OpenUniverse_ESPI}.

The Error Region Counterpart Identifier (ERCI), a core component of \fshort, is based on an enhanced version of VOU-Blazars \citep{VOU-Blazars} combined with custom Python scripts. ERCI is designed to identify potential multi-frequency counterparts within the localization regions of X-ray and \gray\, sources. 
For each 4FGL \gray\, source, within an area about 20\% larger than the 95\% CL signal containment error region listed \citet{4FGL-DR3} catalog,  ERCI retrieves multi-frequency survey data and additional information, such as source variability and spatial extension, from about 90 openly-accessible remote and local catalogs, listed in \autoref{app:survey}. 
The retrieved photometric data points are converted to flux densities and to $\nu F_\nu$ values, corrected for absorption within the Galaxy. These data are then combined to construct a full SED, which is displayed on the front-end for user evaluation and is available for download. 
The source identification  algorithm within ERCI exploits gradients in key regions of the SED, and tests for the presence of non-jet-like components (such as accretion disk, dusty torus and host galaxy), as well as source extension at optical and X-ray energies, to assess the consistency with different source types (blazar, other AGN, clusters of galaxies, Galactic sources etc.). 

\fshort\ also incorporates two tools for estimating the synchrotron peak energy $\nu_{\text{peak}}$ and the flux at the peak $\nu_s F(\nu_s)$ from the SED of candidate blazars: BLAST and W-Peak~\citep{blast,wpeak}. BLAST is a machine-learning-based estimator designed for automated estimation of these parameters directly from the observed SED data points. W-Peak estimates the synchrotron peak frequency and flux by analyzing infrared spectral slopes from WISE and NEOWISE datasets, predictive of blazar jet emission.

Additional tools for investigating source candidates include an Aladin-based multi-wavelength sky map displaying the retrieved catalog data, as well as benchmark SEDs (LSP, ISP, HSP) and host-galaxy or blue-bump templates that can be superimposed on the data for source validation.

\subsection{Workflow for 4FGL Sources} 
\label{subsec:4fgl_workflow}
We conducted an independent search for blazars among the high-Galactic-latitude ($|b| > 10^\circ$) $\gamma$-ray sources of the \lat\ 4FGL catalog~\citep{4FGL-DR3}, with a starting dataset of 5,062~sources. 
We prepared an ASCII file with 4FGL Source name \texttt{Source\_Name}, Right Ascension, Declination \texttt{RAJ2000, DEJ2000}, Long and Short radii of error ellipse at 95\% confidence \texttt{Conf\_95\_SemiMajor, Conf\_95\_SemiMinor}, Position angle (eastward) of the long axis from celestial North \texttt{Conf\_95\_PosAng} and the Name of identified or likely associated source name and class \texttt{ASSOC1, CLASS} along with its position \texttt{RA\_Counterpart, DEC\_Counterpart}.   
From the 4LAC catalog we took the \texttt{ASSOC1, CLASS1} fields. We also annonated the 4FGL association probabilities \texttt{ASSOC\_PROB\_BAY, ASSOC\_PROB\_LR}. We remark that we did not consider the 4FGL secondary association \texttt{ASSOC2, CLASS2} because no position is given. However, we took note of those secondary association for discussion of some specific cases.
This type of information was entered into \fshort\,, which features a special mode for user-input table-data. We run the ERCI tool on all sources one by one. ERCI can provide none, single or multiple candidate counterparts.
Whenever ERCI identified one or more candidates, we verified the validity of each potential counterpart based on its multi-frequency morphological properties (via the Aladin skymaps) and  its SED. Particular attention was given to the SED to ensure the synchrotron peak was consistent with the $\gamma$-ray intensity and spectral slope reported in 4FGL. 
Cases with multiple plausible candidates were resolved by selecting the counterpart with the most compelling SED.  While specific quantitative thresholds for all parameters are complex and depend on the multi-dimensional parameter space explored by ERCI, the final selection of counterparts involved a careful visual inspection of the SEDs by experienced users and, in many cases, by undergraduate and graduate students under expert supervision~(see \autoref{sec:citizen}). The synchrotron peak position estimated by BLAST and W-Peak~\citep{blast,wpeak} provided additional validation of the blazar nature of the counterparts. 

The coincidence with the proposed 4FGL or\footnote{There were cases in which \lat\ sources were associated to blazars in both 4FGL and 4LAC but some only in 4FGL and not in 4LAC and vice-versa. We required a blazar classification in either of the two catalogs.} 4LAC  candidates were also finally scrutinized based on the position (and naming) of the proposed association.
The step-by-step procedure is reported in \autoref{app:step_by_step}.

\subsection{The Role of Flux Variability} \label{subsec:variability}

While the ERCI tool retrieves variability flags from the catalogs that include such information, a detailed quantitative analysis of variability was not the primary driver for counterpart associations.
Multi-wavelength variability was in some cases considered as an additional factor in the evaluation of the counterpart selection process.
For example large variability in the optical or in the infrared band, e.g. from the Zwicky Transient Facility \citep[ZTF,][]{ZTF} or from NEOWISE \citep{NEOWISE} data, which is retrievable directly from \fshort, was used as a confirmation of the blazar nature of a potential counterpart. Cases exhibiting extreme variability in certain bands were also noted and considered in the overall assessment of the counterpart’s nature. Future work may involve a more systematic and quantitative analysis of multi-wavelength variability data to further refine the counterpart associations and blazar classifications.

\section{Results}
\label{sec:results}

\subsection{Numerical Overview}

The results of the search for counterparts to the 5,062 4FGL $\gamma$-ray sources above the Galactic plane ($|b| > 10^\circ$) with \fshort\,, are discussed below 
and are summarized in \autoref{tab:results_classification}. 

\begin{table}[ht]
\centering
\setlength{\tabcolsep}{3pt} 
\begin{tabular}{@{}lrr@{}}  
\hline
\parbox[t]{6cm}{Cases where \fshort\ and 4FGL/4LAC agree} & \textbf{Nr.} & \textbf{\%} \\
\hline
\parbox[t]{6cm}{$\triangleright$ \fshort\ finds same blazar automatically} & 3,380 & 66.8 \\
\parbox[t]{6cm}{$\triangleright$ \fshort\ finds same blazar supervised} & 76 & 1.5 \\
\parbox[t]{6cm}{$\triangleright$ \fshort\ finds same galaxy/radiogalaxy} & 20 & 0.4 \\
\parbox[t]{6cm}{$\triangleright$ \fshort\ finds same pulsar or galactic source} & 163 & 3.2 \\
\parbox[t]{6cm}{$\triangleright$ \fshort\ confirms unassociated} & 854 & 16.9 \\
\hline
\hline
\multicolumn{1}{r}{\textbf{Total agree}} & \textbf{4,493} & \textbf{88.8} \\
\hline
\\[-0.8ex] 
\hline
\parbox[t]{6cm}{Cases where \fshort\ and 4FGL/4LAC disagree} & \textbf{Nr.} & \textbf{\%} \\
\hline
\parbox[t]{6cm}{$\triangleright$ \fshort\ finds a new blazar in previously unassociated source} & 421 & 8.3 \\
\parbox[t]{6cm}{$\triangleright$ \fshort\ finds alternative blazar} & 64 & 1.3 \\
\parbox[t]{6cm}{$\triangleright$ \fshort\ does not find any association} & 49 & 1.0 \\
\parbox[t]{6cm}{$\triangleright$ \fshort\ does not confirm the association} & 16 & 0.3 \\
\parbox[t]{6cm}{$\triangleright$ \fshort\ finds a galaxy instead} & 18 & 0.4 \\
\parbox[t]{6cm}{$\triangleright$ \fshort\ finds a galactic object instead} & 1 & 0.0 \\
\hline
\multicolumn{1}{r}{\textbf{Total disagree}} & \textbf{569} & \textbf{11.2} \\
\hline\hline
\end{tabular}

\caption{Summary of the classification of 4FGL/4LAC sources made with \fshort. 
Percentages are calculated with respect to the total number of 4FGL sources, excluding those associated with Galactic sources in the catalog.
The upper block lists the cases where there is agreement with 4FGL, while the lower block gives the cases where there are disagreements.}
\label{tab:results_classification}
\end{table}

\medskip
We found that \fshort\, agrees with the 4FGL or 4LAC counterparts in 88.8\% of cases (4,493 sources). 
Of these, 3,380 (66.8\%) are blazars confirmed in an automatic way, and 76 (1.5\%) only after supervision\footnote{This latter minority is related to the fact that \fshort\ algorithm does not provide an association in case optical data are missing or on the contrary too numerous. Very likely the \lat\ spatial association with radio sources overcome this problem and find valid candidate in these cases. We are investigating an improvement of our algorithm to address this issue for future release}, for a total of 3,456 blazars.
We also confirm the non-existence of a plausible candidate in 854 (16.9\%) of the cases. The remainder are extragalactic objects (20, 0.4\%, e.g. misaligned jetted and non-jetted AGNs, or near-by galaxies) and Galactic sources (163, 3.2\%, e.g. pulsars, supernova remnants, etc.). 

\medskip
\fshort\, is in disagreement with 4FGL or 4LAC for 569 (11.2\%) \gray\, sources. Among these, the largest population is that of new blazar associations discovered with \fshort\,, which amounts to 421 (8.3\%) objects. In 64 cases (1.4\%) we instead find a different, and more plausible, blazar counterpart. In 49 cases (1.0\%) we cannot confirm a candidate that instead is claimed by 4FGL. In 16 cases (0.3\%), although we observe the presence of a possible counterpart, its nature cannot be  confirmed with high confidence. Finally, \fshort\ finds different counterparts for 19 (0.4\%) sources, including 18 galaxies and 1 
Galactic object.
In the following sections, we discuss individual cases, focusing on blazars, the class of sources most relevant to this work.

\subsection{Agreement between Firmamento and 4FGL}
We focus here on the 4,493 (88.8\%) sources for which there is an agreement between \fshort\, and 4FGL.

\subsubsection{Confirmed blazars}
\label{sec:agree_blazar}
4FGL classifies 3,508 sources (\texttt{CLASS1}) as blazars, differentiating between \texttt{bll, bcu, fsrq} with 47 sources classified \texttt{rdg} as as well as few less populated classes. 
4LAC instead counts 3,383 sources that are also listed in 4FGL. Of these 1,191 are \texttt{bcu}s. Consider that the 4LAC catalog has 3,407 sources in total, 24 of which are not included in 4FGL (listed in \autoref{app:4lac_excluded}), and are not considered in this work. 

Our results agree on the association of blazars with 4FGL sources in 3,456 cases. In 98.5\% of the cases this is achieved automatically by \fshort. In the remaining 1.5\% of the cases (76) sources, \fshort\ did not find the rightful counterpart nor proposed an alternative valid counterpart. This is due to the fact that ERCI tends to give low probability of associations when optical data are either very poor or subject to source confusion. We are currently working on improving this part of the algorithm recognizing that it applies too strict criteria although only for a small minority.

\begin{figure}
    \centering
    \includegraphics[width=0.44\linewidth]{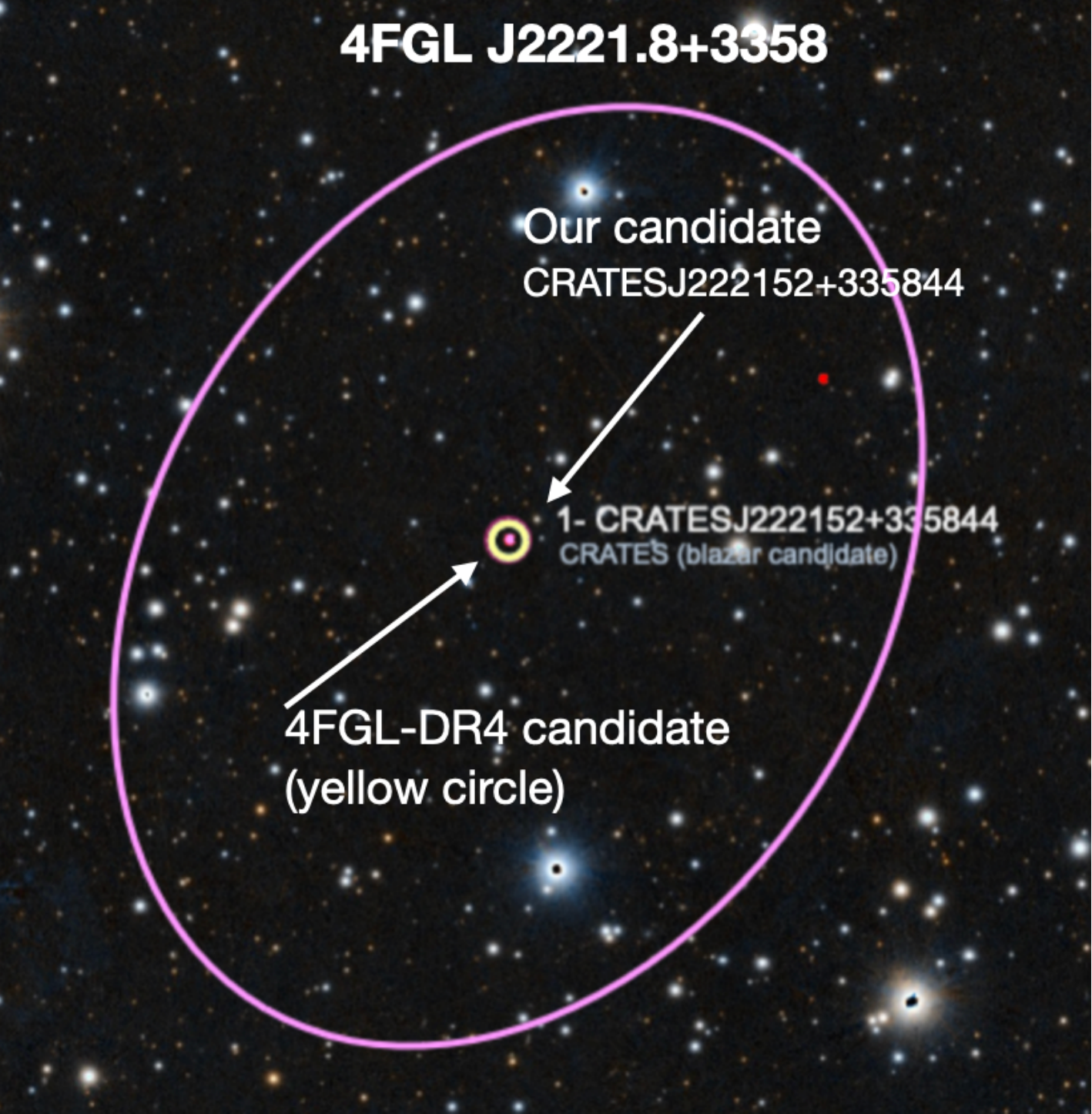}
    \includegraphics[width=0.44\linewidth]{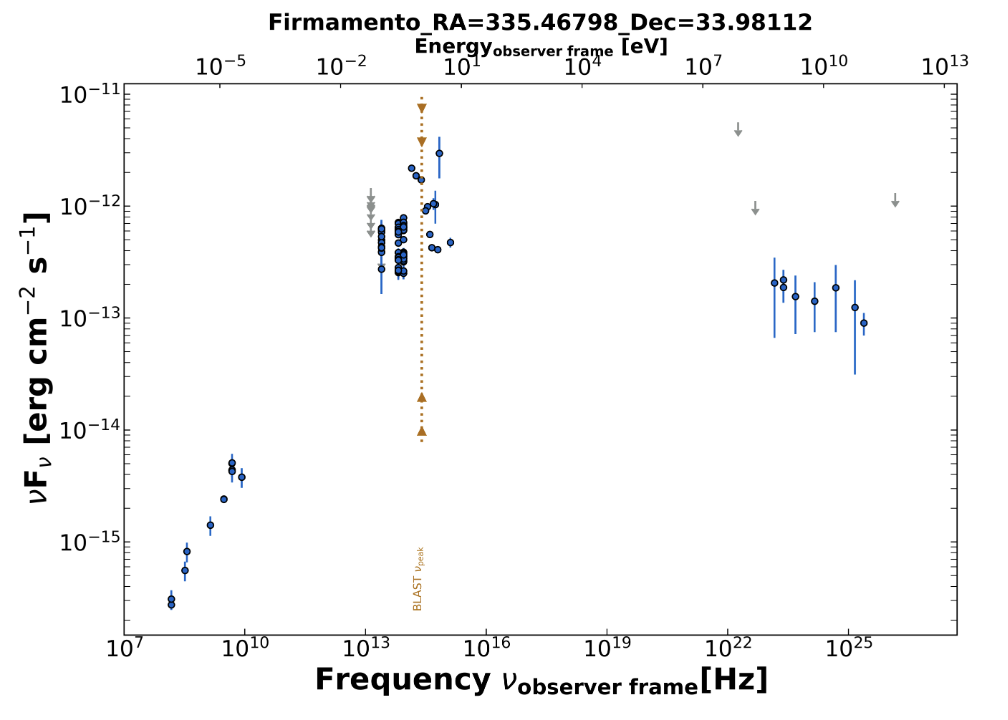}
    \caption{The most common situation exemplified by the case of 4FGL J2221.8+3358 where \fshort\, and 4FGL/4LAC identify the same \gray\, blazar. Left: \fshort\ proposes a single counterpart coincident with the 4LAC counterpart (yellow circle). The purple ellipse in the figure represents the 4FGL error region. Right: The SED of CRATES J222152+3335844, the counterpart of 4FGL J2221.8+3358.}
    \label{fig:yes.blazar} 
\end{figure}

An example of agreement is shown in \autoref{fig:yes.blazar} for the case of 4FGL J2221.8+3358. 
In the figure, the error region of the \gray\, source is shown as a purple ellipse and the 4LAC refined candidate position as a yellow circle. The \fshort\, blazar candidate, which is usually unique, coincides with the 4LAC candidate. The SED of this blazar shows the multi-wavelength spectrum and the position of the synchrotron peak. It is interesting to notice that 213 sources classified as blazars in 4FGL  (5.9\%) were not reported in the 4LAC catalog and vice-versa 5 sources classified as blazar in 4LAC were not similarly classified in 4FGL.

 In \autoref{tab:yesblazar_class} we report the blazar classification obtained with \fshort\ for three \texttt{CLASS} values: \texttt{bll,fsrq,bcu}. One can see that more than half of the 4FGL bll are HSPs, almost all the FSRQs are LSPs while we find that half of the 4FGL \texttt{bcu}s are LSPs. We did not report the statistics for the 4FGL blazars with different classification as their contribution is minor. 

\begin{table}[h!]
\centering
\begin{tabular}{l|ccc}
\hline
$\downarrow$4FGL$\quad$\fshort$\rightarrow$& LSP & ISP & HSP \\
\hline
4FGL \texttt{bll} (1407)  & 304 (22\%) & 335 (24\%) & 768 (55\%) \\
4FGL \texttt{fsrq} (783)  & 763 (97\%) & 12 (2\%)   & 8 (1\%) \\
4FGL \texttt{bcu} (1318)  & 739 (56\%) & 202 (15\%) & 377 (29\%) \\
\hline
TOT (3508)       & 1806 (52\%) & 549 (16\%) & 1153 (33\%) \\
\hline
\end{tabular}
\caption{Distribution of the \fshort\ classification of the 4FGL \texttt{CLASS: bll, fsrq, bcu} blazars also identified by \fshort. }
    \label{tab:yesblazar_class}
    \end{table}

The estimation of the synchrotron peak, not present in 4FGL, is instead provided in 4LAC. We report the comparison of our classification with that of 4LAC in \autoref{sec:sync_peak} over our entire blazar sample.

\subsubsection{Confirmed missing counterparts} 
For 854 4FGL sources (16.9\%), we concur with 4FGL on the absence of a plausible AGN counterpart. This is partly due to the lack of availability of multi-wavelength data at least in one of the bands that characterize a blazar (e.g. X-ray or optical/infrared). In such cases, the \fshort\ algorithm does not reach the threshold for plausible counterpart proposal and return a null association. We remark that if future data will be added, the same algorithm could be run again to re-evaluate the association. 
We remark that among these entries, 14 sources have a non-empty \texttt{4FGL\_CLASS2}, with the following classifications:
\texttt{agn}: 8, \texttt{unk}: 5, \texttt{nlsy1}: 1.

\subsubsection{Confirmed non-blazar counterparts}
We confirm  20 radio galaxies in the sample, associated to extended or double-lobed radio emission, 163 Galactic sources out of which 152 are pulsars. The confirmation is based on both a morphological multi-wavelength visual check with Aladin as well as via SED inspection. We do not investigate further the properties of these non-blazars objects.

\subsection{Disagreements between Firmamento and 4FGL}
We focus here on the 569 (11.2\%) sources for which there is a disagreement between \fshort\ and 4FGL.

\subsection{New Blazar Associations}
\begin{figure}[h!t]
    \centering
    \includegraphics[width=1.0\linewidth]{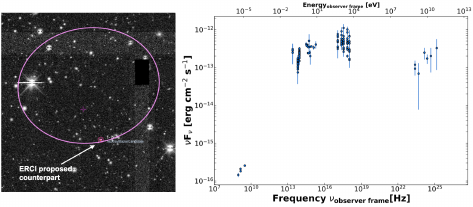}
    \caption{Example of identification of a previously unassociated source (4FGLJ0122.4+1034). Left: the 4FGLJ0122.4+1034 elliptical error region and \fshort\,'s counterpart. Right: the SED of  \fshort\,'s candidate (\fcat\, J012223.6+103213).}
    \label{fig:no.newblazar} 
\end{figure}

Most of the disagreements (74\%) come from the identification of 421 new associations with blazars in previously unassociated  sources. These new findings result from the inclusion of recent multi-wavelength catalogs, the use of independent association algorithms based on multi-wavelength data and not solely on spatial position, and from the visual scrutiny of the SED. An example of a new association is shown in \autoref{fig:no.newblazar} for the case of 4FGLJ0122.4+1034. The candidate association lies at the border of the 4FGL error region. No other counterparts are found by \fshort\,. The SED clearly reveals the blazar nature of this source. In other cases where multiple associations are proposed from \fshort\, each candidate is visually scrutinized. Only in case a single blazar is found, then the association is marked as clear, otherwise as uncertain. This procedure is discussed in \autoref{sec:uncertain}. 

\begin{figure}[h!t]
    \centering
    \includegraphics[width=1\linewidth]{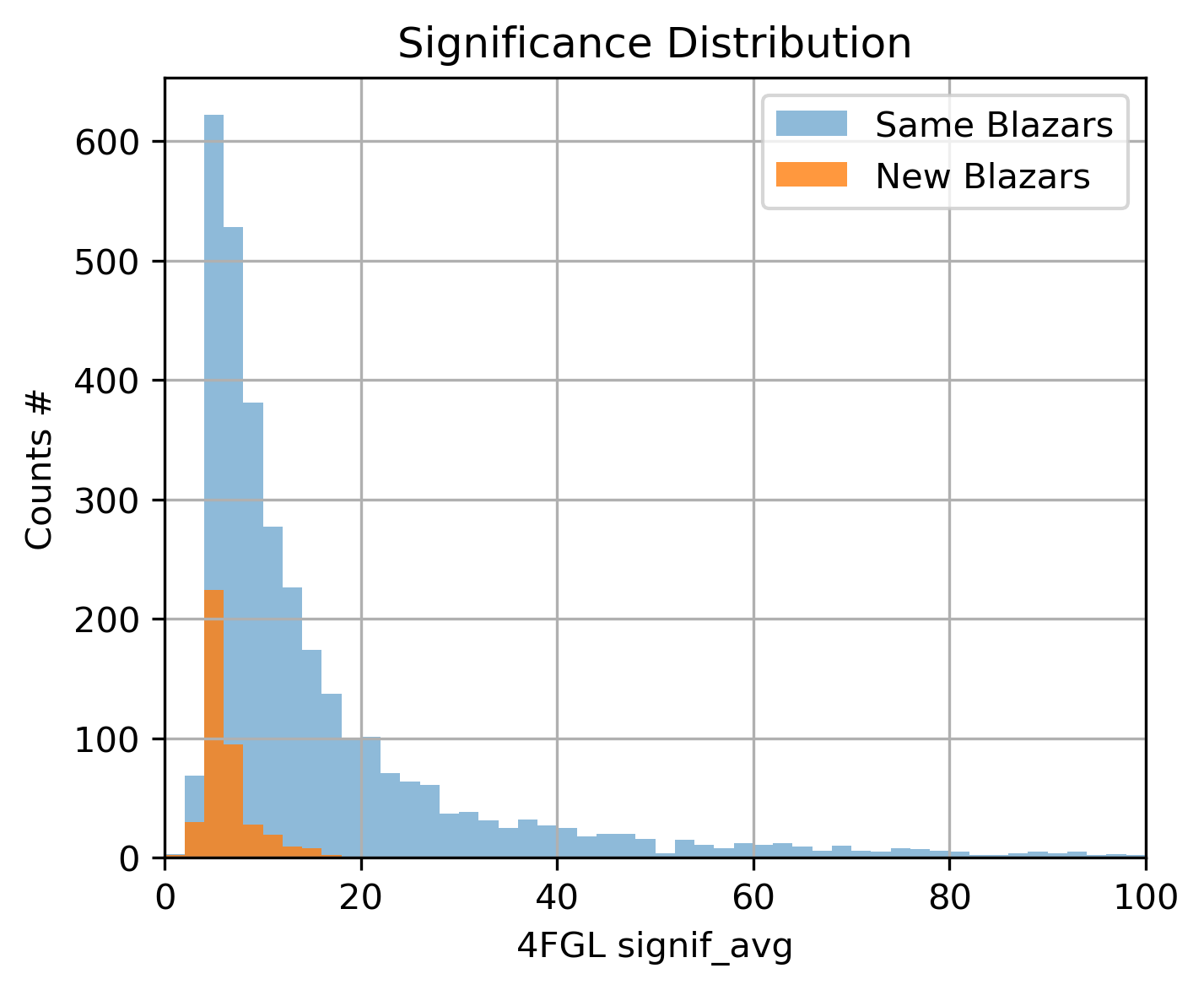}
    \caption{Distribution of the 4FGL \texttt{signif\_avg} significance parameter for the confirmed blazars and the newly discovered.}
    \label{fig:histo_signif_4fgl}
\end{figure}

When comparing with 4FGL we found that a number of these counterparts were classified with \texttt{ASSOC\_PROB\_BAY}$<0.85$ and \texttt{ASSOC\_PROB\_LR}$=0$. For this reason they did not receive a primary association \texttt{ASSOC1}, the 4FGL class designation for associated source, but 21 of them received as \texttt{4FGL\_CLASS2} association, with the following distribution: \texttt{agn} (19), \texttt{sey} (1), \texttt{unk} (1).  Interestingly, the source significance \texttt{Signif\_Avg} distribution for these sources mimics that of the large sample, as shown in \autoref{fig:histo_signif_4fgl}. We checked that 12 out of them coincides with \fshort\ association. For these  \texttt{ASSOC\_PROB\_BAY} ranges from 0.12 to 0.79 while \texttt{ASSOC\_PROB\_LR}$=0$ for all of them. The classification types of the newly discovered blazars are reported in \autoref{tab:newdiffblazar_class}. One can see that the largest fractions are HSP.

\begin{table}[h!t]
    \centering
    \begin{tabular}{c|ccc}
\fshort$\rightarrow$& LSP & ISP & HSP \\
\hline
         New blazar (421) & 130 (31\%) & 96 (23\%) & 195 (46\%) \\
         Different blazar (64) & 17 (27\%) & 8 (12\%) & 39 (61\%) \\
         \hline 
    \end{tabular}
    \caption{\fshort\ classification of the 4FGL unassociated and those for which we find a different blazar.}
    \label{tab:newdiffblazar_class}
\end{table}

\begin{figure}[h!t]
    \centering
    \includegraphics[width=1\linewidth]{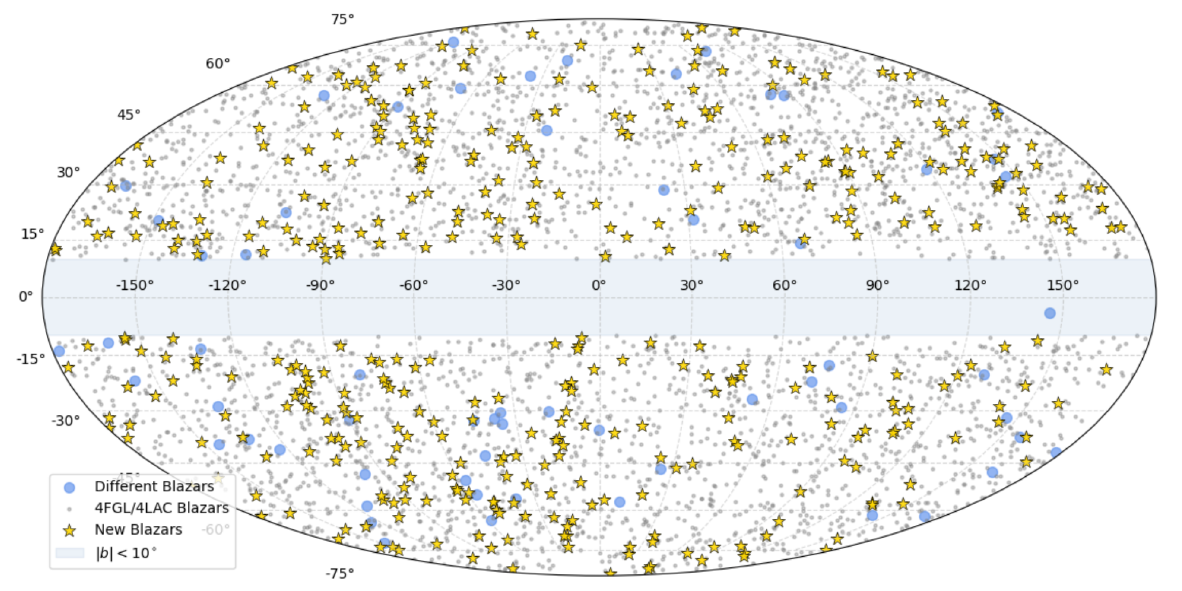}
    \caption{Skymap of 4FGL/4LAC sources (light gray) along with the 421 new blazars discovered (yellow stars symbols) and the sources assigned a different association (light blue circles).}
    \label{fig:1FLAT_skymap}
\end{figure}

A skymap displaying the new \fshort\ associations (yellow stars), and 4FGL confirmed associations (gray points) is shown in \autoref{fig:1FLAT_skymap}. The sky distributions of the two datasets are similar. 

\subsection{Alternative Associations}
\label{sec:alternative_blazar_counterparts}
\begin{figure}[h!t]
    \centering
    \includegraphics[width=1.0\linewidth]{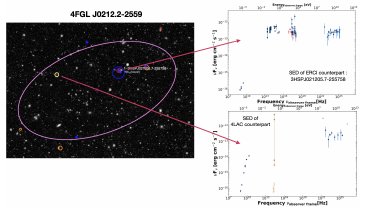}
    \caption{Example of a 4FGL proposed counterpart that is not confirmed by \fshort, which proposed a second blazar. The 4FGL counterpart (lower SED on the right) has a strong radio but has no infrared, optical or X-ray flux, the SED of the 3HSPJ021205.7-255758, the \fshort\ counterpart, clearly fits well with that of an HSP counterpart.}
    \label{fig:no.diffblazar}.
\end{figure}

We identified 64 cases (1.3\% of the sample) where a blazar, different from the one associated in the 4FGL/4LAC catalogs, can be more confidently associated with a $\gamma$-ray source. An example is the case of 4FGL J0212.2-2259, shown in \autoref{fig:no.diffblazar}, where the 4FGL proposed counterpart has an SED with strong radio emission but no detectable infrared or optical flux. In contrast, the counterpart proposed by \fshort\,, 3HSP J021205.7-255758, is an HSP with a well-defined SED that closely matches the flat $\gamma$-ray spectral data from \lat.

Of these sources, in 4FGL, 47 were classified as \texttt{bcu}, 12 a \texttt{bll}, 4 as \texttt{fsrq} and 1 as \texttt{rdg}. A possible reason for this different association may be related to the fact that 4FGL uses catalog association within a spatial range, while \fshort\ uses data correlation and assisted validation. 
The \fshort\ classification of the differently associated blazars is reported in \autoref{tab:newdiffblazar_class}. Also in this case the majority of the sources are HSPs.

\subsection{Non-Confirmation of 4FGL/4LAC Associations}
In 49 other cases (1.0\%),  the 4FGL or 4LAC counterparts  are considered unreliable, and ERCI did not find alternative counterparts within the error region. These source were mostly \texttt{bcu}s in \texttt{4FGL\_CLASS1}.

\subsection{Uncertain classification}
\label{sec:uncertain}

We identified 16 4FGL sources with possible counterparts, but owning to limited data or uncertain SED shapes, 
we feel that their associations are not sufficiently secure. Most of these were again classified as \texttt{bcu}s in \texttt{4FGL\_CLASS1}.

\subsection{The \fcat\, catalog}
Our findings are compiled in a catalog called \fcat\, (first \f\, LAT AGN Table).
The catalog is primarily focused on blazars, although we report some additional results to enable interested scientist to further explore our work. We instead do not list Galactic objects or 
non-active galaxies. In summary the \fcat\, includes:

\begin{itemize}
    \item \textbf{Blazars:} 3,456 blazars found by \fshort\ in agreement with 4FGL; 421 new \fshort\ blazar associations; 64 blazars for which \fshort\ finds a different blazar candidate than 4FGL
    \item \textbf{Uncertain:} 16 sources for which \fshort\ finds an uncertain association 
    \item \textbf{Unassociated:} 854 \gray\, sources for which both \fshort\ and 4FGL do not find an association and 49 sources for which \fshort\ does not find a valid association whereas 4FGL does
    \item \textbf{Radio galaxies:} 20 galaxies found by \fshort\ in agreement with 4FGL and 18 galaxies for which \fshort\ finds a different blazar candidate than 4FGL
\end{itemize}

For all \fshort\ sources we report the 4FGL name, the sky coordinates \texttt{RAJ2000}, \texttt{DEJ2000}, the counterpart class \texttt{CLASS}, the synchrotron peak frequency and flux, \texttt{nu\_syn}, \texttt{nuFnu\_syn} as well as a tag field \texttt{TAG} that reports our internal flags reflecting the level of agreement with 4FGL.

\medskip
In \fcat\, we also report basic information from 4FGL and 4LAC. For 4FGL we report the provenance fields: the name of identified or likely associated primary source \texttt{ASSOC1}, the class designation for associated primary source \texttt{CLASS1}, and corresponding secondary association \texttt{ASSOC2}, \texttt{CLASS2}, the source significance in $\sigma$ units over the 50~MeV to 1~TeV band \texttt{Signif\_Avg}, the photon index when fitting with Power Law \texttt{PL\_Index}, the Energy flux from 100 MeV to 100 GeV obtained by spectral fitting \texttt{Energy\_Flux100}, the Fractional variability \texttt{Frac\_Variability}, the Probability of association according to
the Bayesian method \texttt{ASSOC\_PROB\_BAY}, and the Probability of association according to
the likelihood-ratio method \texttt{ASSOC\_PROB\_LR}. For 4LAC we report the provenance fields \texttt{ASSOC1}, \texttt{CLASS}, the Synchrotron-peak frequency in the observer frame \texttt{nu\_syn} and the $\nu\,f_\nu$ at synchrotron-peak frequency \texttt{nuFnu\_syn}.

We format \fcat\, as a FITS file. The detailed description of the FITS file is reported in \autoref{sec:1FLAT_fits}. A simplified version of the catalog is also available from Firmamento, which gives simple  access to all the multi-frequency data and images.

\subsection{Comparison with Other Association Methods}
\label{sec:comparison}
The identification of 421 new blazar associations among previously unassociated \gray\, sources represents one of the major outcomes of this work.
In addition, we associate 64 sources with counterparts that differ from those reported in 4FGL, highlighting important differences with respect to the 4FGL catalog results.
In the following, we briefly compare our \fshort\,-based technique with the Bayesian and Likelihood Ratio methods adopted by the \lat\ collaboration.

The 4LAC and \fshort\ methods are fundamentally different and depend on markedly different amounts of information. 
As detailed in \autoref{sec:firmamento}, the algorithm implemented in \fshort\ combines very large amount of multi-frequency data with information on source variability and spatial extension. 
\fshort\ determines whether a given error region includes one or more sources that are likely to be blazars or other types of multi-frequency emitters through a two-step process: first it analyzes the shape of the broadband SED of all radio and X-ray sources in the requested area, constructed from approximately 50 catalogs and survey data; then, it builds a more detailed SED from approximately 90 catalogs and spectral databases (see \autoref{app:survey}), which is visually inspected by our team to confirm or reject the candidate(s). 
This final human intervention will eventually be replaced by a machine learning tool trained on the results of this and similar works.

The 4FGL Bayesian and likelihood ratio methods, by contrast, rely solely on single catalogs of previously known sources—namely blazars or flat-spectrum radio sources in the Bayesian case, and radio or X-ray survey catalogs in the likelihood ratio case.

Fig. \ref{fig:no.diffblazar} exemplifies the different outcomes that can result from the application of the 4FGL and \fshort\ methods. 
The 4FGL proposed counterpart (yellow circle and lower SED) was selected both with the Bayesian and the Likelihood ratio methods with association probability of 0.99 and 0.89, respectively. In contrast, the \fshort\ method ignores the 4FGL candidate and selects instead the source named 3HSPJ021205.7-255758, which has a better overall SED. 
There are also cases where the 4FGL methods select reasonable candidates, whereas the \fshort, method fails to identify any. This typically occurs when the candidate is a relatively strong radio source whose SED is of the LSP type, with a very faint optical counterpart and no available X-ray data.
This situation often occurs in regions of the sky where eROSITA survey data are not yet available.

Since the publication of the first 
\lat catalogs, a number of independent teams have also devised alternative methods to aid in the identification of unassociated \gray\, sources.
Some of these approaches use machine learning techniques that rely solely on \gray\, data  \citep{salvetti}, while others combine \gray\, and X-ray data \citep{kaur}.

To support the identification process, \cite{dabrusco} selected a large sample of blazar candidates based on radio data and infrared colors, while \cite{chang20193hsp3HSP} compiled a sample of high-energy-peaked blazars, many of which were expected to be detected by \lat 
and are indeed listed among the confirmed and newly identified blazars in the \fcat\, catalog.
More recent works have focused on the selection of samples of blazars that are detected in the very-high-energy (VHE) band \citep[e.g.][]{1cgh,neronov}. 
A detailed comparison between our work and these efforts — some of which are considered in the 4FGL-DR4 and 4LAC-DR3 papers — is beyond the scope of this study. 

\section{Analysis and Discussion}
\label{sec:analysis}
The main result of this work is the discovery of 421 new  blazar identifications in the 4FGL catalog. In the following we characterize this population properties.

\subsection{\gray\, flux}
\begin{figure}[h!t]
\includegraphics[width=\columnwidth]{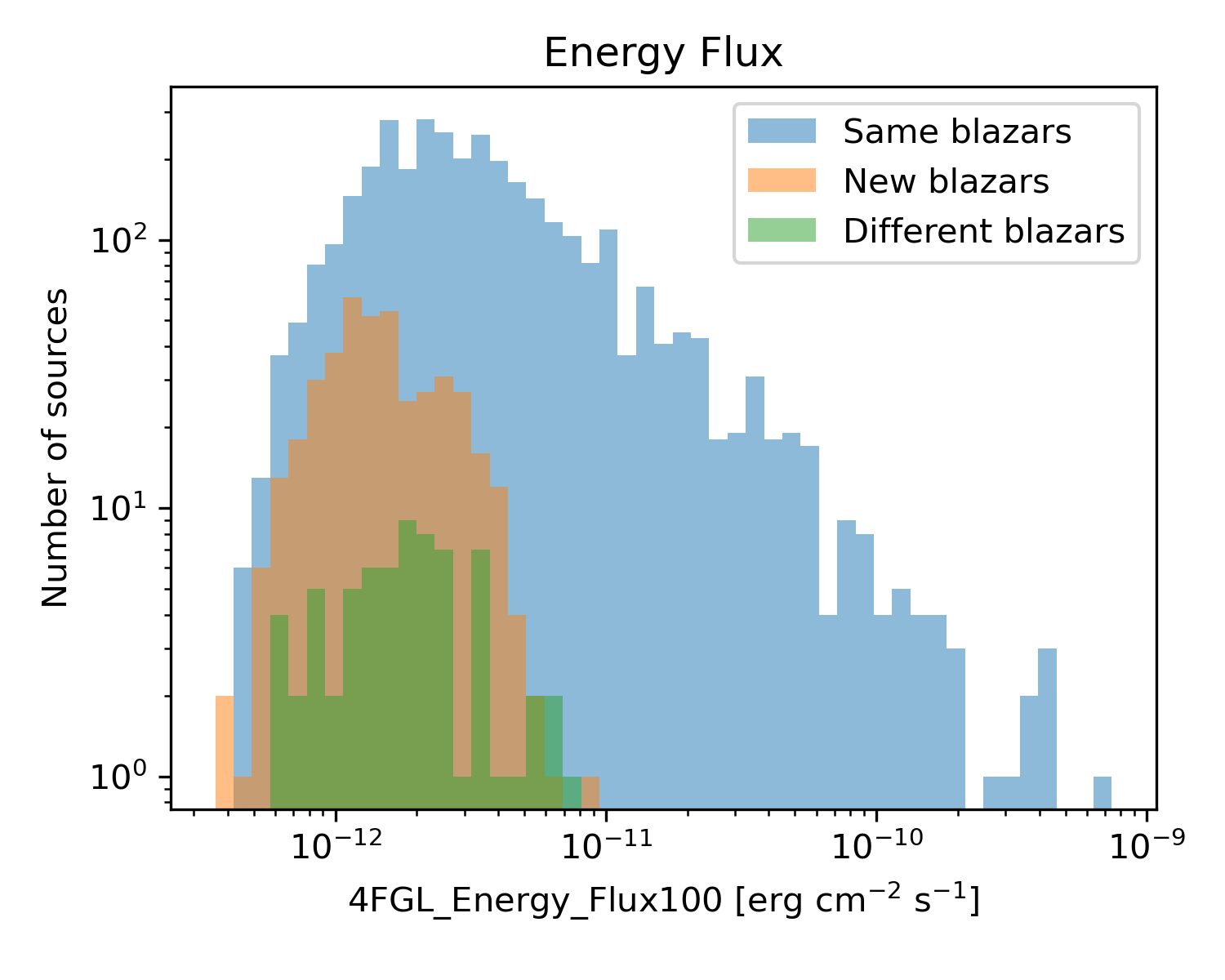}
    \caption{The distribution of the flux \texttt{Energy\_Flux100} for the same, different and new blazars.}
    \label{fig:flux_distr} 
\end{figure}

In \autoref{fig:flux_distr} we show the distribution of \gray\, fluxes as reported in 4FGL, \texttt{Energy\_Flux100} field, for the confirmed blazars, the newly discovered and those for which we have a different classification. The confirmed blazars have a wide distribution of fluxes, reaching values as high as $10^{-9}$~erg cm$^{-2}$ s$^{-1}$. The newly discovered blazars and the different associations are instead less bright, with maximum flux of the order of $10^{-11}$~erg cm$^{-2}$ s$^{-1}$. This distribution likely reflects that these are fainter blazars and are therefore more difficult to detect and reliably associate.

\subsection{Synchrotron peak frequency and blazar classes}
\label{sec:sync_peak}
In this work we classify blazars as LSP if $\nu_{\text{peak}} < 10^{13.5}$ Hz, ISP if $10^{13.5} \leq \nu_{\text{peak}} < 10^{15}$ Hz and HSP if $\nu_{\text{peak}} \geq 10^{15}$ Hz \citep{giommipadovani}. This is slightly different from the 4LAC catalog,  which uses $10^{14}$~Hz to separate LSP/ISP. 
As mentioned, we estimate the synchrotron peak with two independent algorithms, \texttt{BLAST} \citep{blast} and \texttt{wpeak} ~\citep{wpeak}. While BLAST always returns an estimation of $\nu_{\text{peak}}$, \texttt{wpeak} only provides an estimate if the flux from the host galaxy is negligible. This happens roughly half of the cases. The two estimations are consistent and compatible, with null mean difference and 5\% RMSD, which suggests that the two methods are robust and their estimations accurate. 

\begin{figure}[h!t]
\includegraphics[width=\columnwidth]{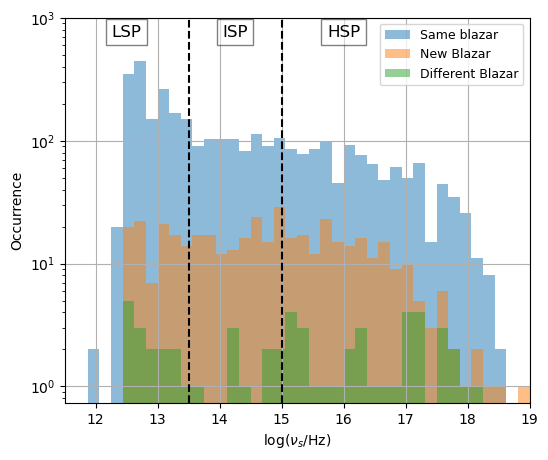}
    \caption{The distribution of the synchrotron peak frequency in the sample. Altogether, we show a total blazar population of 1,696 (43.0\%) LSP, 925 (23.5\%) ISP and 1,322 (33.5\%) HSP.}
    \label{fig:nupeak_distr} 
\end{figure}

The distribution of synchrotron peak frequencies in the \fcat\, catalog is plotted in \autoref{fig:nupeak_distr} for the newly discovered blazar, those with alternative associations, and for the confirmed blazars. The overall distribution shows LSPs as the most common type, consistent with the LAT team catalogs, see \autoref{tab:blazar_class} for details. 

\begin{table}[htbp]
\centering
\caption{\label{tab:blazar_class}Classification summary for blazars. "Same" are those that both 4FGL and \fshort\ find. "New" are those found only in \fshort.~"Different" are the alternative blazar \fshort\ finds with respect to 4FGL.}
\begin{tabular}{l||c|c|c||c}
\toprule
 & \textbf{Same} & \textbf{New} & \textbf{Different} & \textbf{All} \\
\midrule
LSP & 1824 -- 52.8\% & 130 -- 30.9\% & 17 -- 26.6\% & 50.0\% \\
ISP & 560 -- 16.2\%  & 96 -- 22.8\%  & 8 -- 12.5\%  & 16.8\% \\
HSP & 1072 -- 31.0\% & 195 -- 46.3\% & 39 -- 60.9\% & 33.1\% \\
\midrule
Tot & 3456           & 421           & 64            & 3941     \\
\bottomrule
\end{tabular}
\end{table}

The newly associated blazars exhibit a greater prevalence of ISPs and HSPs compared to the general population. This is very likely due to the fact that HSPs are typically found with low significance in 4FGL due to the position of the high-energy SED bump that is located at higher energies and to a low value of the Compton dominance, i.e. the ratio between the high energy and synchrotron peak. The mean values of \texttt{4FGL\_ASSOC\_PROB\_BAY} for the three blazar classes are 0.78, 0.82, and 0.92 for HSP, ISP, and LSP, respectively.
On average, LSPs have the highest association probability, with a tighter spread, while HSPs have the lowest mean and the widest range.

\begin{figure}[h!t]
	\includegraphics[width=\columnwidth]{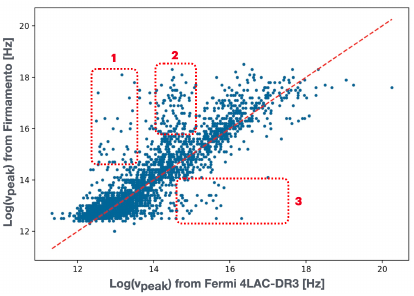}
    \caption{The Log($\nu_{\text{peak}}$) from the \lat\, 4LAC catalog is plotted against the Log($\nu_{\text{peak}}$) estimated with \texttt{BLAST} within \fshort. The diagonal red dashed line represents equal values. The two estimates generally agree within less than one decade, however there are also large differences, highlighted by the dotted box areas labeled 1, 2 and 3 (see text for details).}
\label{fig:nupeaks_comparison}
\end{figure}

\autoref{fig:nupeaks_comparison} compares \fcat\, Log($\nu_{\text{peak}}$) values with those estimated in the 4LAC catalog, for blazars where this parameter could be estimated in both datasets; the red diagonal line indicates equal values. 
The two independent estimations generally cluster around the red line, with some scatter, likely due to differences in the SED datasets available in \fshort\ compared to those used in 4LAC, as well as the different estimation methods. In \fshort, the estimation method is homogeneous, relying  on a machine learning  
approach or a tool based on an algorithm, and thus largely independent of human intervention.
We note that in \autoref{fig:nupeaks_comparison} there are also regions, highlighted by the dotted rectangles  labeled 1, 2, and 3, where significant disagreement between 4LAC and \fshort\ is observed. 
We examined the cases where \texttt{BLAST} estimated a much larger  $\nu_{\text{peak}}$ value compared to 4LAC (areas labeled 1 and 2). In all instances, the higher $\nu_{\text{peak}}$ 
values were attributed to the availability of high-quality X-ray measurements, which may not have been available when the 4LAC catalog was compiled.
The limited X-ray data available to 4LAC and the potential misinterpretation of X-ray data as the end of the synchrotron component, rather than as IC emission in \fshort\ could explain the discrepancies observed in the area labeled 3. 

\subsection{Spectral properties}
\begin{figure}[h!t]
\includegraphics[width=\columnwidth]{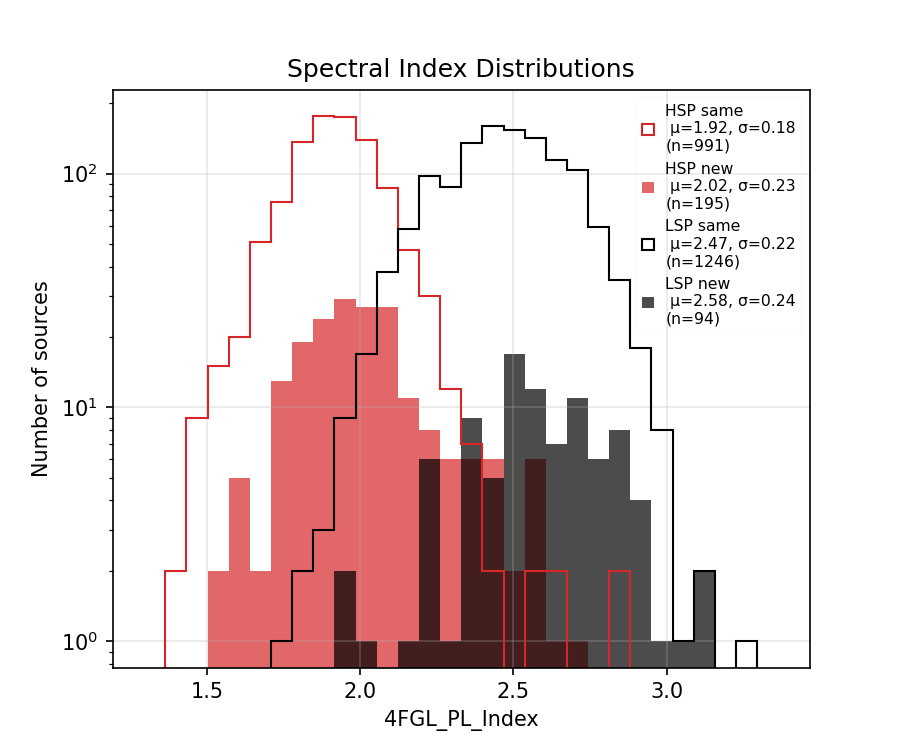}
\caption{
 The distribution of 4LAC \gray\, spectral slopes in the subsamples of LSP and HSP blazars, including both confirmed and newly associated sources. Since nearly all newly associated blazars have a lower \gray\, flux than the confirmed blazars, we restrict the comparison to sources with \texttt{Energy\_Flux100} $<10^{-11}$ erg cm$^{-2}$ s$^{-1}$, for consistency.
}
\label{fig:slopeshistogram}
\end{figure}

To further compare the sample of new associations with the confirmed associations in 4FGL, we plot in \autoref{fig:slopeshistogram} the histogram of \gray\, spectral indices for the HSP and LSP blazar subsamples.
The histograms show both the large sample of confirmed blazars and the newly discovered blazars, along with some relevant statistics.
The distributions of \gray\, spectral indices of the new blazars appear similar to that of the confirmed blazars. Therefore we conclude that our new associations are an extension of the same type of blazars found in 4FGL rather than a new population with peculiar properties.

\begin{figure}[h!t]
\includegraphics[width=\columnwidth]{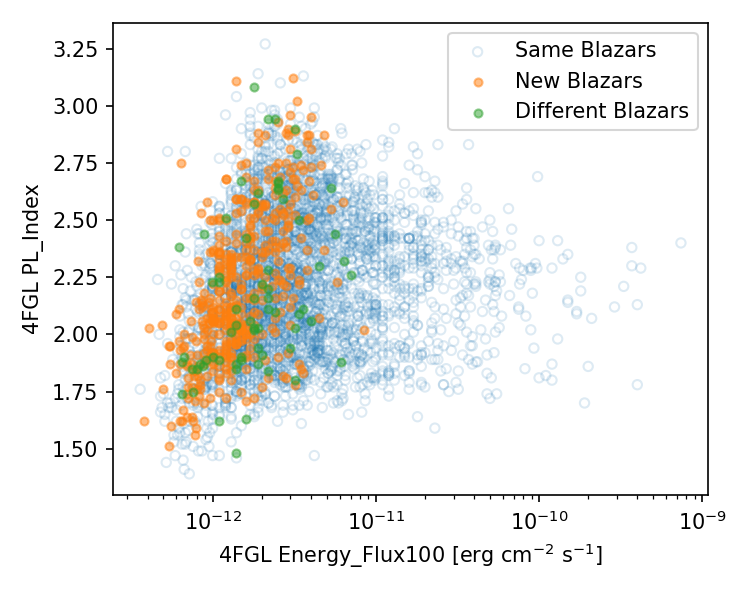}
    \caption{The power-law photon index from the 4FGL catalog is plotted against the \gray\, flux, for the cases of the confirmed associations, newly identified blazars, and with different associations. The new and different identifications are spread over the spectral index but are largely confined to \gray\, fluxes lower than $\approx 10^{-11}$ erg cm$^{-2}$ s$^{-1}$. }
\label{fig:slopevsflux}
\end{figure}

\autoref{fig:slopevsflux} shows the 4FGL photon index versus the 4FGL \gray\, flux for the confirmed, new and differently associated blazars. Again, the new population of blazars overlaps well with the general population, in the lower flux range.

\section{Educational Engagement}
\label{sec:citizen}
The \fshort\ platform was designed with a strong commitment to inclusivity and educational engagement, aligning with the principles of the Open Universe (OU) initiative~\citep{OpenUniverse_ESPI}. This commitment facilitated the active participation of both graduate and undergraduate students in the \fcat\, project. These students, some with limited or no prior experience in blazar research, played a relevant role in the analysis of \lat\ $\gamma-$ray sources.

The activity started with a preliminary project within the Italian Ministry of Education PCTO program\footnote{\url{https://www.istruzione.it/pon/avviso\_pcto.html}}.
Undergraduate students participated in the research through a series of online sessions led by experts in the field, spanning approximately six months. These sessions focused on training them in the functionalities of the \fshort\ platform and the procedures for blazar discovery. This training finished with the discovery of an initial sample of 54 potential blazars among a subset of 4FGL sources. This was presented in conferences and international gatherings, and received significant media coverage and recognition, 
and obtained a prize in the FAST 2023 contest\footnote{See e.g.\ \textit{NYUAD Citizen Researcher} (2023), 
\url{https://www.instagram.com/reel/Cq-X1XaAuQY}; 
\textit{Firmamento Workshop} (2023), 
\url{https://nyuad.nyu.edu/en/events/2023/april/firmamento-workshop.html}; 
\textit{La Nuova Venezia} (2022), 
\url{https://l.infn.it/1bo}; 
\textit{INFN News} (2023), 
\url{https://l.infn.it/1bn}.}.
Ultimately it resulted  in the peer-reviewed proceedings \cite{Fronte_2023}. \fshort\ received important upgrades after that experience to improve the algorithm and the usage. We have cross-checked that now 45 out of 54 new blazars proposed in \citep{Fronte_2023} are also found in this work, while 9 are confirmed without associations: 4FGL J0152.9-1109, J0944.6+5729, J1409.8+7921, J1504.6+4343, J1519.7+6727, J1658.5+4315, J1706.4+6428, J2030.3-5038, J2237.8+2430, due to \f\, improvement.

The process of identifying blazar in this work  was extended to multiple participants, including the students that were in the pilot project. About half of the authors of this manuscript are undergraduate and they scrutinized about 38\% of the sources.
The classifications made by the participants were subsequently reviewed by experts in the field to ensure the accuracy and reliability of the results. This was done at random on sources that the students classified with certainty, while it was done systematically on the sources that the students labeled for further checks. These were typically sources with multiple proposed counterparts.

\section{Discussion and Conclusions}
\label{sec:conclusion}
In this study, we have presented the \fcat\,, an independently derived catalog of blazar counterparts of high-Galactic latitude \gray\,  sources in the \lat\ 4FGL catalog, constructed using the \fshort\, web-based platform. Our results showed a high level of agreement (over 88\%) with the associations reported in the 4FGL and 4LAC catalogs, but also significant differences.
A key finding of this work is the identification of 421 new credible blazar associations for previously unassociated $\gamma$-ray sources in the 4FGL catalog. This significantly reduces the fraction of unidentified high-Galactic latitude \lat\ sources from approximately 25\% to 17\%. We also identified 64 alternative blazar counterparts and found 49 cases where the 4FGL/4LAC association was not confirmed.

The \fcat\, catalog, with its refined and expanded list of AGN counterparts of \lat\ $\gamma$-ray sources, has several important astrophysical implications. The increased number of identified blazars, particularly HSPs and ISPs among the previously unassociated sources, contributes to a more complete understanding of the $\gamma$-ray emitting blazar population. The detailed SED information and synchrotron peak parameter estimates provided in the catalog are valuable for detailed population studies 
and help modeling the physical processes within relativistic jets.

However, the methodology employed in this work also has limitations. While \fshort\ leverages a vast amount of multi-frequency data and sophisticated tools, the identification of counterparts may still be challenging in large error regions, crowded fields, or when multi-wavelength data are sparse or of poor quality. Also, the reliance on visual inspection of SEDs, while allowing for expert judgment, can introduce a degree of subjectivity.

Potential biases in the new associations should also be considered. For instance, if our method is more sensitive to blazars with specific multi-wavelength properties, this could lead to a biased representation of the overall unassociated source population.

Future studies could address these limitations and further enhance the \fcat\, catalog.
Incorporating quantitative variability measures in a more systematic way could improve the robustness of the process.
In addition, developing more sophisticated machine-learning techniques for automated SED classification and counterpart association would further enhance its efficiency.
The upcoming release of the full eROSITA X-ray survey data is expected to significantly aid in identifying counterparts for currently unassociated $\gamma$-ray sources, providing an opportunity to further validate and expand the \fcat\, catalog.

Finally, we note that a systematic analysis of redshift distributions was not attempted in this work, 
as reliable spectroscopic information is only available for a subset of the sources. 
Since the main focus here is on counterpart associations, we defer a comprehensive redshift study to a future dedicated publication.

\medskip
In summary, the \fcat catalog, available online through the \fshort\, platform, provides a complementary and independently derived list of AGN associated with LAT \gray\, sources.
It is intended as a useful reference for very-high-energy \gray\, observations, multi-wavelength campaigns, and studies of \gray –emitting AGN, particularly in view of the significant number of new associations presented here.

\bigskip
\section*{Data Availability}
\label{sec:data_availability}

 All the data used in this paper is public and available via the \fshort\ and other platforms.

\section*{Acknowledgements}
\label{sec:acknowledgements}
\begin{small}
PG expresses his gratitude to the Center for Astrophysics and Space Science (CASS) of New York University, Abu Dhabi, for supporting his research visits at NYU-Abu Dhabi. MD acknowledges support from Prof. Alice Scelsi and Prof. Alberto Signoretti for the guidance of the students of Liceo Scientifico Statale Ugo Morin who participated in this work in the framework of the PCTO exchange between the high school and the University of Padova. UBA acknowledges the financial support of the Ministry of Science, Technology and Innovation of Brazil to the Open Universe Initiative, through which his participation in this work was funded. MG acknowledges support from Coordenação de Aperfeiçoamento de Pessoal de Nível Superior – Brasil (CAPES) – Finance Code 001 and Conselho Nacional de Desenvolvimento Científico e Tecnológico - Brasil (CNPq). EP  acknowledges funding for the project “SKYNET: Deep Learning for Astroparticle Physics”, PRIN 2022 (CUP: D53D23002610006). NS acknowledges the support by the Higher Education and Science Committee of the Republic of Armenia, in the frames of the research project No 23LCG-1C004.
\end{small}

\bibliography{biblio,biblio_survey}{}

\begin{thebibliography}{}
\expandafter\ifx\csname natexlab\endcsname\relax\def\natexlab#1{#1}\fi
\providecommand{\url}[1]{\href{#1}{#1}}
\providecommand{\dodoi}[1]{doi:~\href{http://doi.org/#1}{\nolinkurl{#1}}}
\providecommand{\doeprint}[1]{\href{http://ascl.net/#1}{\nolinkurl{http://ascl.net/#1}}}
\providecommand{\doarXiv}[1]{\href{https://arxiv.org/abs/#1}{\nolinkurl{https://arxiv.org/abs/#1}}}

\bibitem[{A. {Abdo} {et~al.}(2010){Abdo}, {Ackermann}, {Agudo}, {Ajello}, {Aller}, {Aller}, {Angelakis}, {Arkharov}, {Axelsson}, {Bach}, {Baldini}, {Ballet}, {Barbiellini}, {Bastieri}, {Baughman}, {Bechtol}, {Bellazzini}, {Benitez}, {Berdyugin}, {Berenji}, {Blandford}, {Bloom}, {Boettcher}, {Bonamente}, {Borgland}, {Bregeon}, {Brez}, {Brigida}, {Bruel}, {Burnett}, {Burrows}, {Buson}, {Caliandro}, {Calzoletti}, {Cameron}, {Capalbi}, {Caraveo}, {Carosati}, {Casandjian}, {Cavazzuti}, {Cecchi}, {{\c{C}}elik}, {Charles}, {Chaty}, {Chekhtman}, {Chen}, {Chiang}, {Chincarini}, {Ciprini}, {Claus}, {Cohen-Tanugi}, {Colafrancesco}, {Cominsky}, {Conrad}, {Costamante}, {Cutini}, {D'ammando}, {Deitrick}, {D'Elia}, {Dermer}, {de Angelis}, {de Palma}, {Digel}, {Donnarumma}, {Silva}, {Drell}, {Dubois}, {Dultzin}, {Dumora}, {Falcone}, {Farnier}, {Favuzzi}, {Fegan}, {Focke}, {Forn{\'e}}, {Fortin}, {Frailis}, {Fuhrmann}, {Fukazawa}, {Funk}, {Fusco}, {G{\'o}mez}, {Gargano}, {Gasparrini}, {Gehrels}, {Germani}, {Giebels},
  {Giglietto}, {Giommi}, {Giordano}, {Giuliani}, {Glanzman}, {Godfrey}, {Grenier}, {Gronwall}, {Grove}, {Guillemot}, {Guiriec}, {Gurwell}, {Hadasch}, {Hanabata}, {Harding}, {Hayashida}, {Hays}, {Healey}, {Heidt}, {Hiriart}, {Horan}, {Hoversten}, {Hughes}, {Itoh}, {Jackson}, {J{\'o}hannesson}, {Johnson}, {Johnson}, {Jorstad}, {Kadler}, {Kamae}, {Katagiri}, {Kataoka}, {Kawai}, {Kennea}, {Kerr}, {Kimeridze}, {Kn{\"o}dlseder}, {Kocian}, {Kopatskaya}, {Koptelova}, {Konstantinova}, {Kovalev}, {Kovalev}, {Kurtanidze}, {Kuss}, {Lande}, {Larionov}, {Latronico}, {Leto}, {Lindfors}, {Longo}, {Loparco}, {Lott}, {Lovellette}, {Lubrano}, {Madejski}, {Makeev}, {Marchegiani}, {Marscher}, {Marshall}, {Max-Moerbeck}, {Mazziotta}, {McConville}, {McEnery}, {Meurer}, {Michelson}, {Mitthumsiri}, {Mizuno}, {Moiseev}, {Monte}, {Monzani}, {Morselli}, {Moskalenko}, {Murgia}, {Nestoras}, {Nilsson}, {Nizhelsky}, {Nolan}, {Norris}, {Nuss}, {Ohsugi}, {Ojha}, {Omodei}, {Orlando}, {Ormes}, {Osborne}, {Ozaki}, {Pacciani}, {Padovani},
  {Pagani}, {Page}, {Paneque}, {Panetta}, {Parent}, {Pasanen}, {Pavlidou}, {Pelassa}, {Pepe}, {Perri}, {Pesce-Rollins}, {Piranomonte}, {Piron}, {Pittori}, {Porter}, {Puccetti}, {Rahoui}, {Rain{\`o}}, {Raiteri}, {Rando}, {Razzano}, {Reimer}, \& {Reimer}}]{2010Abdo}
{Abdo}, A., {Ackermann}, M., {Agudo}, I., {et~al.} 2010, \bibinfo{title}{{The Spectral Energy Distribution of Fermi Bright Blazars},} \apj, 716, 30, \dodoi{10.1088/0004-637X/716/1/30}

\bibitem[{A.~A. Abdo {et~al.}(2013)Abdo {et~al.}}]{abdo2013secondF2PSR}
Abdo, A.~A., {et~al.} 2013, \bibinfo{title}{The second Fermi Large Area Telescope catalog of gamma-ray pulsars,} ApJS, 208, 17

\bibitem[{S. Abdollahi {et~al.}(2020)Abdollahi, Acero, Ackermann, Ajello, Atwood, Axelsson, Baldini, Ballet, Barbiellini, Bastieri, Becerra~Gonzalez, Bellazzini, Berretta, Bissaldi, Blandford, Bloom, Bonino, Bottacini, Brandt, Bregeon, Bruel, Buehler, Burnett, Buson, Cameron, Caputo, Caraveo, Casandjian, Castro, Cavazzuti, Charles, Chaty, Chen, Cheung, Chiaro, Ciprini, Cohen-Tanugi, Cominsky, Coronado-Blázquez, Costantin, Cuoco, Cutini, D’Ammando, DeKlotz, Torre~Luque, de~Palma, Desai, Digel, Lalla, Mauro, Venere, Domínguez, Dumora, Dirirsa, Fegan, Ferrara, Franckowiak, Fukazawa, Funk, Fusco, Gargano, Gasparrini, Giglietto, Giommi, Giordano, Giroletti, Glanzman, Green, Grenier, Griffin, Grondin, Grove, Guiriec, Harding, Hayashi, Hays, Hewitt, Horan, Jóhannesson, Johnson, Kamae, Kerr, Kocevski, Kovac’evic’, Kuss, Landriu, Larsson, Latronico, Lemoine-Goumard, Li, Liodakis, Longo, Loparco, Lott, Lovellette, Lubrano, Madejski, Maldera, Malyshev, Manfreda, Marchesini, Marcotulli, Martí-Devesa, Martin,
  Massaro, Mazziotta, McEnery, Mereu, Meyer, Michelson, Mirabal, Mizuno, Monzani, Morselli, Moskalenko, Negro, Nuss, Ojha, Omodei, Orienti, Orlando, Ormes, Palatiello, Paliya, Paneque, Pei, Peña-Herazo, Perkins, Persic, Pesce-Rollins, Petrosian, Petrov, Piron, Poon, Porter, Principe, Rainò, Rando, Razzano, Razzaque, Reimer, Reimer, Remy, Reposeur, Romani, Parkinson, Schinzel, Serini, Sgrò, Siskind, Smith, Spandre, Spinelli, Strong, Suson, Tajima, Takahashi, Tak, Thayer, Thompson, Tibaldo, Torres, Torresi, Valverde, Klaveren, Zyl, Wood, Yassine, \& Zaharijas}]{4FGL-DR1}
Abdollahi, S., Acero, F., Ackermann, M., {et~al.} 2020, \bibinfo{title}{Fermi Large Area Telescope Fourth Source Catalog,} The Astrophysical Journal Supplement Series, 247, 33, \dodoi{10.3847/1538-4365/ab6bcb}

\bibitem[{S. {Abdollahi} {et~al.}(2022){Abdollahi}, {Acero}, {Baldini}, {Ballet}, {Bastieri}, {Bellazzini}, {Berenji}, {Berretta}, {Bissaldi}, {Blandford}, {Bloom}, {Bonino}, {Brill}, {Britto}, {Bruel}, {Burnett}, {Buson}, {Cameron}, {Caputo}, {Caraveo}, {Castro}, {Chaty}, {Cheung}, {Chiaro}, {Cibrario}, {Ciprini}, {Coronado-Bl{\'a}zquez}, {Crnogorcevic}, {Cutini}, {D'Ammando}, {De Gaetano}, {Digel}, {Di Lalla}, {Dirirsa}, {Di Venere}, {Dom{\'\i}nguez}, {Fallah Ramazani}, {Fegan}, {Ferrara}, {Fiori}, {Fleischhack}, {Franckowiak}, {Fukazawa}, {Funk}, {Fusco}, {Galanti}, {Gammaldi}, {Gargano}, {Garrappa}, {Gasparrini}, {Giacchino}, {Giglietto}, {Giordano}, {Giroletti}, {Glanzman}, {Green}, {Grenier}, {Grondin}, {Guillemot}, {Guiriec}, {Gustafsson}, {Harding}, {Hays}, {Hewitt}, {Horan}, {Hou}, {J{\'o}hannesson}, {Karwin}, {Kayanoki}, {Kerr}, {Kuss}, {Landriu}, {Larsson}, {Latronico}, {Lemoine-Goumard}, {Li}, {Liodakis}, {Longo}, {Loparco}, {Lott}, {Lubrano}, {Maldera}, {Malyshev}, {Manfreda},
  {Mart{\'\i}-Devesa}, {Mazziotta}, {Mereu}, {Meyer}, {Michelson}, {Mirabal}, {Mitthumsiri}, {Mizuno}, {Moiseev}, {Monzani}, {Morselli}, {Moskalenko}, {Negro}, {Nuss}, {Omodei}, {Orienti}, {Orlando}, {Paneque}, {Pei}, {Perkins}, {Persic}, {Pesce-Rollins}, {Petrosian}, {Pillera}, {Poon}, {Porter}, {Principe}, {Rain{\`o}}, {Rando}, {Rani}, {Razzano}, {Razzaque}, {Reimer}, {Reimer}, {Reposeur}, {S{\'a}nchez-Conde}, {Saz Parkinson}, {Scotton}, {Serini}, {Sgr{\`o}}, {Siskind}, {Smith}, {Spandre}, {Spinelli}, {Sueoka}, {Suson}, {Tajima}, {Tak}, {Thayer}, {Thompson}, {Torres}, {Troja}, {Valverde}, {Wood}, \& {Zaharijas}}]{4FGL-DR3}
{Abdollahi}, S., {Acero}, F., {Baldini}, L., {et~al.} 2022, \bibinfo{title}{{Incremental Fermi Large Area Telescope Fourth Source Catalog},} \apjs, 260, 53, \dodoi{10.3847/1538-4365/ac6751}

\bibitem[{G.~O. Abell {et~al.}(1989)Abell {et~al.}}]{abell1989}
Abell, G.~O., {et~al.} 1989, \bibinfo{title}{A catalog of rich clusters of galaxies,} ApJS, 70, 1

\bibitem[{B. Abolfathi {et~al.}(2018)Abolfathi {et~al.}}]{abolfathi2018fourteenthSDSS}
Abolfathi, B., {et~al.} 2018, \bibinfo{title}{The fourteenth data release of the Sloan digital sky survey: First spectroscopic data from the extended baryon oscillation spectroscopic survey and from the second phase of the Apache Point Observatory Galactic Evolution Experiment,} ApJS, 235, 42

\bibitem[{M. Ajello {et~al.}(2017)Ajello {et~al.}}]{ajello20173fhl3FHL}
Ajello, M., {et~al.} 2017, \bibinfo{title}{3FHL: The third catalog of hard Fermi-LAT sources,} ApJS, 232, 18

\bibitem[{M. {Ajello} {et~al.}(2022){Ajello}, {Baldini}, {Ballet}, {Bastieri}, {Becerra Gonzalez}, {Bellazzini}, {Berretta}, {Bissaldi}, {Bonino}, {Brill}, {Bruel}, {Buson}, {Caputo}, {Caraveo}, {Cheung}, {Chiaro}, {Cibrario}, {Ciprini}, {Crnogorcevic}, {Cutini}, {D'Ammando}, {De Gaetano}, {Di Lalla}, {Di Venere}, {Dom{\'\i}nguez}, {Ramazani}, {Ferrara}, {Fiori}, {Fukazawa}, {Funk}, {Fusco}, {Gammaldi}, {Gargano}, {Garrappa}, {Gasparrini}, {Giglietto}, {Giordano}, {Giroletti}, {Green}, {Grenier}, {Guiriec}, {Horan}, {Hou}, {Kayanoki}, {Kuss}, {Larsson}, {Latronico}, {Lewis}, {Li}, {Liodakis}, {Longo}, {Loparco}, {Lott}, {Lovellette}, {Lubrano}, {Madejski}, {Maldera}, {Manfreda}, {Mart{\'\i}-Devesa}, {Mazziotta}, {Mereu}, {Michelson}, {Mirabal}, {Mitthumsiri}, {Mizuno}, {Monzani}, {Morselli}, {Moskalenko}, {Negro}, {Ojha}, {Orienti}, {Orlando}, {Ormes}, {Pei}, {Pe{\~n}a-Herazo}, {Persic}, {Pesce-Rollins}, {Petrosian}, {Pillera}, {Poon}, {Porter}, {Principe}, {Rain{\`o}}, {Rando}, {Rani}, {Razzano}, {Razzaque},
  {Reimer}, {Reimer}, {Scotton}, {Serini}, {Sgr{\`o}}, {Siskind}, {Spandre}, {Spinelli}, {Suson}, {Tajima}, {Torres}, {Valverde}, {Yassin}, \& {Zaharijas}}]{4LAC-DR3}
{Ajello}, M., {Baldini}, L., {Ballet}, J., {et~al.} 2022, \bibinfo{title}{{The Fourth Catalog of Active Galactic Nuclei Detected by the Fermi Large Area Telescope: Data Release 3},} \apjs, 263, 24, \dodoi{10.3847/1538-4365/ac9523}

\bibitem[{B. {Arsioli} {et~al.}(2025){Arsioli}, {Chang}, \& {Ighina}}]{1cgh}
{Arsioli}, B., {Chang}, Y.-L., \& {Ighina}, L. 2025, \bibinfo{title}{{Mapping the cosmic gamma-ray horizon: the 1CGH catalogue of Fermi-LAT detections above 10 GeV},} \mnras, 539, 1458, \dodoi{10.1093/mnras/staf329}

\bibitem[{B. {Arsioli} {et~al.}(2020){Arsioli}, {Chang}, \& {Musiimenta}}]{2bigb}
{Arsioli}, B., {Chang}, Y.~L., \& {Musiimenta}, B. 2020, \bibinfo{title}{{Extreme and high synchrotron peak blazars beyond 4FGL: The 2BIGB {\ensuremath{\gamma}}-ray catalogue},} \mnras, 493, 2438, \dodoi{10.1093/mnras/staa368}

\bibitem[{A. Avakyan {et~al.}(2023)Avakyan {et~al.}}]{avakyan2023xrbcatsXRBCAT}
Avakyan, A., {et~al.} 2023, \bibinfo{title}{XRBcats: Galactic low-mass X-ray binary catalogue,} A\&A, 675, A199

\bibitem[{J. Ballet {et~al.}(2023)Ballet, Bruel, Burnett, \& Lott}]{4FGL-DR4}
Ballet, J., Bruel, P., Burnett, T.~H., \& Lott, B. 2023, \bibinfo{title}{{Fermi Large Area Telescope Fourth Source Catalog Data Release 4 (4FGL-DR4)},} \doarXiv{2307.12546}

\bibitem[{V. Beckmann \& C. Shrader(2012)Beckmann \& Shrader}]{beckmann2012agn}
Beckmann, V., \& Shrader, C. 2012, Active Galactic Nuclei (Wiley-VCH).
\newblock \url{https://www.wiley.com/en-au/Active+Galactic+Nuclei-p-9783527666805}

\bibitem[{M.~R. Blanton {et~al.}(2017)Blanton {et~al.}}]{blanton2017sloanSDSS}
Blanton, M.~R., {et~al.} 2017, \bibinfo{title}{Sloan digital sky survey IV: Mapping the Milky Way, nearby galaxies, and the distant universe,} AJ, 154, 28

\bibitem[{M. {Bonato} {et~al.}(2019){Bonato}, {Liuzzo}, {Herranz}, {Gonz{\'a}lez-Nuevo}, {Bonavera}, {Tucci}, {Massardi}, {De Zotti}, {Negrello}, \& {Zwaan}}]{alma}
{Bonato}, M., {Liuzzo}, E., {Herranz}, D., {et~al.} 2019, \bibinfo{title}{{ALMA photometry of extragalactic radio sources},} \mnras, 485, 1188, \dodoi{10.1093/mnras/stz465}

\bibitem[{E. {Bonning} {et~al.}(2012){Bonning}, {Urry}, {Bailyn}, {Buxton}, {Chatterjee}, {Coppi}, {Fossati}, {Isler}, \& {Maraschi}}]{smarts}
{Bonning}, E., {Urry}, C.~M., {Bailyn}, C., {et~al.} 2012, \bibinfo{title}{{SMARTS Optical and Infrared Monitoring of 12 Gamma-Ray Bright Blazars},} \apj, 756, 13, \dodoi{10.1088/0004-637X/756/1/13}

\bibitem[{H. Brunner {et~al.}(2022)Brunner {et~al.}}]{brunner2022erositaEFEDS}
Brunner, H., {et~al.} 2022, \bibinfo{title}{The eROSITA Final Equatorial Depth Survey (eFEDS)-X-ray catalogue,} A\&A, 661, A1

\bibitem[{A. Bulgarelli {et~al.}(2019)Bulgarelli {et~al.}}]{bulgarelli2019vizier2AGILE}
Bulgarelli, A., {et~al.} 2019, \bibinfo{title}{VizieR Online Data Catalog: Second AGILE catalogue of gamma-ray sources (Bulgarelli+, 2019),} VizieR Online Data Catalog

\bibitem[{J.~R. {Callingham} {et~al.}(2017){Callingham}, {Ekers}, {Gaensler}, {Line}, {Hurley-Walker}, {Sadler}, {Tingay}, {Hancock}, {Bell}, {Dwarakanath}, {For}, {Franzen}, {Hindson}, {Johnston-Hollitt}, {Kapi{\'n}ska}, {Lenc}, {McKinley}, {Morgan}, {Offringa}, {Procopio}, {Staveley-Smith}, {Wayth}, {Wu}, \& {Zheng}}]{eprs}
{Callingham}, J.~R., {Ekers}, R.~D., {Gaensler}, B.~M., {et~al.} 2017, \bibinfo{title}{{Extragalactic Peaked-spectrum Radio Sources at Low Frequencies},} \apj, 836, 174, \dodoi{10.3847/1538-4357/836/2/174}

\bibitem[{Y.~L. {Chang} {et~al.}(2020){Chang}, {Brandt}, \& {Giommi}}]{VOU-Blazars}
{Chang}, Y.~L., {Brandt}, C.~H., \& {Giommi}, P. 2020, \bibinfo{title}{{The Open Universe VOU-Blazars tool},} Astronomy and Computing, 30, 100350, \dodoi{10.1016/j.ascom.2019.100350}

\bibitem[{Y.-L. Chang {et~al.}(2019)Chang {et~al.}}]{chang20193hsp3HSP}
Chang, Y.-L., {et~al.} 2019, \bibinfo{title}{The 3HSP catalogue of extreme and high-synchrotron peaked blazars,} A\&A, 632, A77

\bibitem[{P. Collaboration(2016)Collaboration}]{planck2016}
Collaboration, P. 2016, \bibinfo{title}{Planck 2015 results. XXVII. The Second Planck Catalogue of Sunyaev-Zeldovich Sources,} A\&A, 594, A27

\bibitem[{J.~J. Condon {et~al.}(1998)Condon {et~al.}}]{condon1998nraoNVSS}
Condon, J.~J., {et~al.} 1998, \bibinfo{title}{The NRAO VLA sky survey,} AJ, 115, 1693

\bibitem[{R.~M. Cutri {et~al.}(2012)Cutri {et~al.}}]{cutri2012vizierWISE-VIZIER}
Cutri, R.~M., {et~al.} 2012, \bibinfo{title}{VizieR online data catalog: WISE all-sky data release (Cutri, 2012),} VizieR Online Data Catalog, II

\bibitem[{R. {D'Abrusco} {et~al.}(2019){D'Abrusco}, {{\'A}lvarez Crespo}, {Massaro}, {Campana}, {Chavushyan}, {Landoni}, {La Franca}, {Masetti}, {Milisavljevic}, {Paggi}, {Ricci}, \& {Smith}}]{dabrusco}
{D'Abrusco}, R., {{\'A}lvarez Crespo}, N., {Massaro}, F., {et~al.} 2019, \bibinfo{title}{{Two New Catalogs of Blazar Candidates in the WISE Infrared Sky},} \apjs, 242, 4, \dodoi{10.3847/1538-4365/ab16f4}

\bibitem[{C. {De Breuck} {et~al.}(2002){De Breuck}, {Tang}, {de Bruyn}, {R{\"o}ttgering}, \& {van Breugel}}]{wish}
{De Breuck}, C., {Tang}, Y., {de Bruyn}, A.~G., {R{\"o}ttgering}, H., \& {van Breugel}, W. 2002, \bibinfo{title}{{A sample of ultra steep spectrum sources selected from the Westerbork In the Southern Hemisphere (WISH) survey},} \aap, 394, 59, \dodoi{10.1051/0004-6361:20021115}

\bibitem[{C.~D. Dermer \& G. Menon(2009)Dermer \& Menon}]{dermer2009high}
Dermer, C.~D., \& Menon, G. 2009, High Energy Radiation from Black Holes: Gamma Rays, Cosmic Rays, and Neutrinos (Princeton University Press)

\bibitem[{M. Doro {et~al.}(2021)Doro, Nigro, Prandini, Tramacere, Delfino, Delgado, do~Souto, Jouvin, \& Rico}]{Doro:2019zvt}
Doro, M., Nigro, C., Prandini, E., {et~al.} 2021, \bibinfo{title}{{Toward a Public MAGIC Gamma-Ray Telescope Legacy Data Portal},} PoS, ICRC2019, 666, \dodoi{10.22323/1.358.0666}

\bibitem[{J.~N. {Douglas} {et~al.}(1996){Douglas}, {Bash}, {Bozyan}, {Torrence}, \& {Wolfe}}]{texas}
{Douglas}, J.~N., {Bash}, F.~N., {Bozyan}, F.~A., {Torrence}, G.~W., \& {Wolfe}, C. 1996, \bibinfo{title}{{The Texas Survey of Radio Sources Covering -35.5 degrees < declination < 71.5 degrees at 365 MHz},} \aj, 111, 1945, \dodoi{10.1086/117932}

\bibitem[{S.~W. Duchesne {et~al.}(2024)Duchesne {et~al.}}]{duchesne2024rapidRACSMID}
Duchesne, S.~W., {et~al.} 2024, \bibinfo{title}{The Rapid ASKAP Continuum Survey V: Cataloguing the sky at 1367.5 MHz and the second data release of RACS-mid,} PASA, 41, e003

\bibitem[{P.~R.~M. {Eisenhardt} {et~al.}(2020){Eisenhardt}, {Marocco}, {Fowler}, {Meisner}, {Kirkpatrick}, {Garcia}, {Jarrett}, {Koontz}, {Marchese}, {Stanford}, {Caselden}, {Cushing}, {Cutri}, {Faherty}, {Gelino}, {Gonzalez}, {Mainzer}, {Mobasher}, {Schlegel}, {Stern}, {Teplitz}, \& {Wright}}]{CatWISE}
{Eisenhardt}, P. R.~M., {Marocco}, F., {Fowler}, J.~W., {et~al.} 2020, \bibinfo{title}{{The CatWISE Preliminary Catalog: Motions from WISE and NEOWISE Data},} \apjs, 247, 69, \dodoi{10.3847/1538-4365/ab7f2a}

\bibitem[{M. Elvis {et~al.}(1992)Elvis {et~al.}}]{elvis1992}
Elvis, M., {et~al.} 1992, \bibinfo{title}{The Einstein IPC Slew Survey,} ApJS, 80, 257

\bibitem[{I.~N. Evans {et~al.}(2010)Evans {et~al.}}]{evans2010}
Evans, I.~N., {et~al.} 2010, \bibinfo{title}{The Chandra Source Catalog,} ApJS, 189, 37

\bibitem[{P.~A. Evans {et~al.}(2020)Evans {et~al.}}]{evans2020}
Evans, P.~A., {et~al.} 2020, \bibinfo{title}{2SXPS: Swift-XRT point source catalog,} ApJS, 247, 54

\bibitem[{W.~B. {Everett} {et~al.}(2020){Everett}, {Zhang}, {Crawford}, {Vieira}, {Aravena}, {Archipley}, {Austermann}, {Benson}, {Bleem}, {Carlstrom}, {Chang}, {Chapman}, {Crites}, {de Haan}, {Dobbs}, {George}, {Halverson}, {Harrington}, {Holder}, {Holzapfel}, {Hrubes}, {Knox}, {Lee}, {Luong-Van}, {Mangian}, {Marrone}, {McMahon}, {Meyer}, {Mocanu}, {Mohr}, {Natoli}, {Padin}, {Pryke}, {Reichardt}, {Reuter}, {Ruhl}, {Sayre}, {Schaffer}, {Shirokoff}, {Spilker}, {Stalder}, {Staniszewski}, {Stark}, {Story}, {Switzer}, {Vanderlinde}, {Wei{\ss}}, \& {Williamson}}]{sptsz}
{Everett}, W.~B., {Zhang}, L., {Crawford}, T.~M., {et~al.} 2020, \bibinfo{title}{{Millimeter-wave Point Sources from the 2500 Square Degree SPT-SZ Survey: Catalog and Population Statistics},} \apj, 900, 55, \dodoi{10.3847/1538-4357/ab9df7}

\bibitem[{E.~W. Flesch(2015)Flesch}]{flesch2015}
Flesch, E.~W. 2015, \bibinfo{title}{The Million Quasars (Milliquas) Catalog, Version 8,} PASA, 32, e010

\bibitem[{L. Fronte {et~al.}(2023)Fronte, Mazzon, Metruccio, Munaretto, Doro, Giommi, Viale, \& de~Almeida}]{Fronte_2023}
Fronte, L., Mazzon, B., Metruccio, F., {et~al.} 2023, \bibinfo{title}{A catalog of new Blazar candidates with Open Universe by High School students,} Journal of Physics: Conference Series, 2429, 012045, \dodoi{10.1088/1742-6596/2429/1/012045}

\bibitem[{ {Gaia Collaboration} {et~al.}(2023){Gaia Collaboration}, {Vallenari}, {Brown}, {Prusti}, {de Bruijne}, {Arenou}, {Babusiaux}, {Biermann}, {Creevey}, {Ducourant}, {Evans}, {Eyer}, {Guerra}, {Hutton}, {Jordi}, {Klioner}, {Lammers}, {Lindegren}, {Luri}, {Mignard}, {Panem}, {Pourbaix}, {Randich}, {Sartoretti}, {Soubiran}, {Tanga}, {Walton}, {Bailer-Jones}, {Bastian}, {Drimmel}, {Jansen}, {Katz}, {Lattanzi}, {van Leeuwen}, {Bakker}, {Cacciari}, {Casta{\~n}eda}, {De Angeli}, {Fabricius}, {Fouesneau}, {Fr{\'e}mat}, {Galluccio}, {Guerrier}, {Heiter}, {Masana}, {Messineo}, {Mowlavi}, {Nicolas}, {Nienartowicz}, {Pailler}, {Panuzzo}, {Riclet}, {Roux}, {Seabroke}, {Sordo}, {Th{\'e}venin}, {Gracia-Abril}, {Portell}, {Teyssier}, {Altmann}, {Andrae}, {Audard}, {Bellas-Velidis}, {Benson}, {Berthier}, {Blomme}, {Burgess}, {Busonero}, {Busso}, {C{\'a}novas}, {Carry}, {Cellino}, {Cheek}, {Clementini}, {Damerdji}, {Davidson}, {de Teodoro}, {Nu{\~n}ez Campos}, {Delchambre}, {Dell'Oro}, {Esquej},
  {Fern{\'a}ndez-Hern{\'a}ndez}, {Fraile}, {Garabato}, {Garc{\'\i}a-Lario}, {Gosset}, {Haigron}, {Halbwachs}, {Hambly}, {Harrison}, {Hern{\'a}ndez}, {Hestroffer}, {Hodgkin}, {Holl}, {Jan{\ss}en}, {Jevardat de Fombelle}, {Jordan}, {Krone-Martins}, {Lanzafame}, {L{\"o}ffler}, {Marchal}, {Marrese}, {Moitinho}, {Muinonen}, {Osborne}, {Pancino}, {Pauwels}, {Recio-Blanco}, {Reyl{\'e}}, {Riello}, {Rimoldini}, {Roegiers}, {Rybizki}, {Sarro}, {Siopis}, {Smith}, {Sozzetti}, {Utrilla}, {van Leeuwen}, {Abbas}, {{\'A}brah{\'a}m}, {Abreu Aramburu}, {Aerts}, {Aguado}, {Ajaj}, {Aldea-Montero}, {Altavilla}, {{\'A}lvarez}, {Alves}, {Anders}, {Anderson}, {Anglada Varela}, {Antoja}, {Baines}, {Baker}, {Balaguer-N{\'u}{\~n}ez}, {Balbinot}, {Balog}, {Barache}, {Barbato}, {Barros}, {Barstow}, {Bartolom{\'e}}, {Bassilana}, {Bauchet}, {Becciani}, {Bellazzini}, {Berihuete}, {Bernet}, {Bertone}, {Bianchi}, {Binnenfeld}, {Blanco-Cuaresma}, {Blazere}, {Boch}, {Bombrun}, {Bossini}, {Bouquillon}, {Bragaglia}, {Bramante}, {Breedt},
  {Bressan}, {Brouillet}, {Brugaletta}, {Bucciarelli}, {Burlacu}, {Butkevich}, {Buzzi}, {Caffau}, {Cancelliere}, {Cantat-Gaudin}, {Carballo}, {Carlucci}, {Carnerero}, {Carrasco}, {Casamiquela}, {Castellani}, {Castro-Ginard}, {Chaoul}, {Charlot}, {Chemin}, {Chiaramida}, {Chiavassa}, {Chornay}, {Comoretto}, {Contursi}, {Cooper}, {Cornez}, {Cowell}, {Crifo}, {Cropper}, {Crosta}, {Crowley}, {Dafonte}, {Dapergolas}, {David}, {David}, {de Laverny}, {De Luise}, \& {De March}}]{gaiadr3}
{Gaia Collaboration}, {Vallenari}, A., {Brown}, A.~G.~A., {et~al.} 2023, \bibinfo{title}{{Gaia Data Release 3. Summary of the content and survey properties},} \aap, 674, A1, \dodoi{10.1051/0004-6361/202243940}

\bibitem[{P. Giommi(2025)Giommi}]{Giommi2025}
Giommi, P. 2025, in Proceedings of the 8th Heidelberg \gray\, symposium, Milan, Vol. Accepted for publication.
\newblock \url{https://arxiv.org/abs/2503.04434}

\bibitem[{P. {Giommi} {et~al.}(2002){Giommi}, {Capalbi}, {Fiocchi}, {Memola}, {Perri}, {Piranomonte}, {Rebecchi}, \& {Massaro}}]{bepposax}
{Giommi}, P., {Capalbi}, M., {Fiocchi}, M., {et~al.} 2002, in Blazar Astrophysics with BeppoSAX and Other Observatories, ed. P.~{Giommi}, E.~{Massaro}, \& G.~{Palumbo}, 63, \dodoi{10.48550/arXiv.astro-ph/0209596}

\bibitem[{P. {Giommi} \& P. {Padovani}(2021){Giommi} \& {Padovani}}]{giommipadovani}
{Giommi}, P., \& {Padovani}, P. 2021, \bibinfo{title}{{Astrophysical Neutrinos and Blazars},} Universe, 7, 492, \dodoi{10.3390/universe7120492}

\bibitem[{P. Giommi {et~al.}(2024{\natexlab{a}})Giommi, Sahakyan, Israyelyan, \& Manvelyan}]{wpeak}
Giommi, P., Sahakyan, N., Israyelyan, D., \& Manvelyan, M. 2024{\natexlab{a}}, \bibinfo{title}{{The Remarkable Predictive Power of Infrared Data of Blazars},} Astrophys. J., 963, 48, \dodoi{10.3847/1538-4357/ad20cb}

\bibitem[{P. Giommi {et~al.}(2018)Giommi, Arrigo, Barres~de Almeida, de~Angelis, del Rio~Vera, di~Ciaccio, di~Pippo, Iacovoni, \& Pollock}]{OpenUniverse_ESPI}
Giommi, P., Arrigo, G., Barres~de Almeida, U., {et~al.} 2018, in ESPI-UNISPACE+50, Vol. Space 2030 and Space 4.0: Synergies for Capacity Building in the XXI Century.
\newblock \url{https://arxiv.org/abs/1805.08505}

\bibitem[{P. Giommi {et~al.}(2024{\natexlab{b}})Giommi {et~al.}}]{giommi_prep}
Giommi, P., {et~al.} 2024{\natexlab{b}}, \bibinfo{title}{1OUSX catalog,} in preparation

\bibitem[{T. {Glauch} {et~al.}(2022){Glauch}, {Kerscher}, \& {Giommi}}]{blast}
{Glauch}, T., {Kerscher}, T., \& {Giommi}, P. 2022, \bibinfo{title}{{BlaST - A machine-learning estimator for the synchrotron peak of blazars},} Astronomy and Computing, 41, 100646, \dodoi{10.1016/j.ascom.2022.100646}

\bibitem[{Y.~A. Gordon {et~al.}(2020)Gordon, Boyce, O’Dea, Rudnick, Andernach, Vantyghem, Baum, Bui, \& Dionyssiou}]{gordon2020catalogVLASSQL}
Gordon, Y.~A., Boyce, M.~M., O’Dea, C.~P., {et~al.} 2020, \bibinfo{title}{A catalog of very large array sky survey epoch 1 quick look components, sources, and host identifications,} Research Notes of the AAS, 4, 175

\bibitem[{D.~A. Green(2025)Green}]{green2025updatedSNRGREEN}
Green, D.~A. 2025, \bibinfo{title}{An updated catalogue of 310 Galactic supernova remnants and their statistical properties,} J. Astrophys. Astron., 46, 1

\bibitem[{P.~C. {Gregory} \& J.~J. {Condon}(1991){Gregory} \& {Condon}}]{GB87}
{Gregory}, P.~C., \& {Condon}, J.~J. 1991, \bibinfo{title}{{The 87GB Catalog of Radio Sources Covering 0 degrees < delta < +75 degrees at 4.85 GHz},} \apjs, 75, 1011, \dodoi{10.1086/191559}

\bibitem[{P.~C. {Gregory} {et~al.}(1996){Gregory}, {Scott}, {Douglas}, \& {Condon}}]{GB6}
{Gregory}, P.~C., {Scott}, W.~K., {Douglas}, K., \& {Condon}, J.~J. 1996, \bibinfo{title}{{The GB6 Catalog of Radio Sources},} \apjs, 103, 427, \dodoi{10.1086/192282}

\bibitem[{C.~L. Hale {et~al.}(2021)Hale {et~al.}}]{hale2021rapidRACS}
Hale, C.~L., {et~al.} 2021, \bibinfo{title}{The rapid ASKAP continuum survey paper II: First stokes I Source catalogue data release,} arXiv preprint arXiv:2109.00956

\bibitem[{D.~E. Harris {et~al.}(1990)Harris {et~al.}}]{harris1990einsteinIPC2E}
Harris, D.~E., {et~al.} 1990, \bibinfo{title}{The Einstein Observatory catalog of IPC X-ray sources,} Cambridge

\bibitem[{S.~E. Healey {et~al.}(2007)Healey {et~al.}}]{healey2007cratesCRATES}
Healey, S.~E., {et~al.} 2007, \bibinfo{title}{CRATES: an all-sky survey of flat-spectrum radio sources,} ApJS, 171, 61

\bibitem[{D.~J. Helfand {et~al.}(2015)Helfand, White, \& Becker}]{helfand2015lastFIRST}
Helfand, D.~J., White, R.~L., \& Becker, R.~H. 2015, \bibinfo{title}{The last of FIRST: the final catalog and source identifications,} ApJ, 801, 26

\bibitem[{G. {Helou} \& D.~W. {Walker}(1988){Helou} \& {Walker}}]{iras}
{Helou}, G., \& {Walker}, D.~W., eds. 1988, {Infrared Astronomical Satellite (IRAS) Catalogs and Atlases.Volume 7: The Small Scale Structure Catalog.}, Vol.~7

\bibitem[{N. {Hurley-Walker} {et~al.}(2017){Hurley-Walker}, {Callingham}, {Hancock}, {Franzen}, {Hindson}, {Kapi{\'n}ska}, {Morgan}, {Offringa}, {Wayth}, {Wu}, {Zheng}, {Murphy}, {Bell}, {Dwarakanath}, {For}, {Gaensler}, {Johnston-Hollitt}, {Lenc}, {Procopio}, {Staveley-Smith}, {Ekers}, {Bowman}, {Briggs}, {Cappallo}, {Deshpande}, {Greenhill}, {Hazelton}, {Kaplan}, {Lonsdale}, {McWhirter}, {Mitchell}, {Morales}, {Morgan}, {Oberoi}, {Ord}, {Prabu}, {Shankar}, {Srivani}, {Subrahmanyan}, {Tingay}, {Webster}, {Williams}, \& {Williams}}]{gleam}
{Hurley-Walker}, N., {Callingham}, J.~R., {Hancock}, P.~J., {et~al.} 2017, \bibinfo{title}{{GaLactic and Extragalactic All-sky Murchison Widefield Array (GLEAM) survey - I. A low-frequency extragalactic catalogue},} \mnras, 464, 1146, \dodoi{10.1093/mnras/stw2337}

\bibitem[{H.~T. {Intema} {et~al.}(2017){Intema}, {Jagannathan}, {Mooley}, \& {Frail}}]{tgss}
{Intema}, H.~T., {Jagannathan}, P., {Mooley}, K.~P., \& {Frail}, D.~A. 2017, \bibinfo{title}{{The GMRT 150 MHz all-sky radio survey. First alternative data release TGSS ADR1},} \aap, 598, A78, \dodoi{10.1051/0004-6361/201628536}

\bibitem[{D.~H. Jones {et~al.}(2009)Jones {et~al.}}]{jones2009}
Jones, D.~H., {et~al.} 2009, \bibinfo{title}{The 6dF Galaxy Survey,} MNRAS, 399, 683

\bibitem[{A. {Kaur} {et~al.}(2019){Kaur}, {Falcone}, {Stroh}, {Kennea}, \& {Ferrara}}]{kaur}
{Kaur}, A., {Falcone}, A.~D., {Stroh}, M.~D., {Kennea}, J.~A., \& {Ferrara}, E.~C. 2019, \bibinfo{title}{{Classification of New X-Ray Counterparts for Fermi Unassociated Gamma-Ray Sources Using the Swift X-Ray Telescope},} \apj, 887, 18, \dodoi{10.3847/1538-4357/ab4ceb}

\bibitem[{N.~V. Kharchenko {et~al.}(2013)Kharchenko {et~al.}}]{kharchenko2013globalMWSC}
Kharchenko, N.~V., {et~al.} 2013, \bibinfo{title}{Global survey of star clusters in the Milky Way-II. The catalogue of basic parameters,} A\&A, 558, A53

\bibitem[{J. Kube {et~al.}(2003)Kube {et~al.}}]{kube2003cvcatCVCAT}
Kube, J., {et~al.} 2003, \bibinfo{title}{CVcat: An interactive database on cataclysmic variables,} A\&A, 404, 1159

\bibitem[{H. {Kuehr} {et~al.}(1981){Kuehr}, {Witzel}, {Pauliny-Toth}, \& {Nauber}}]{kuehr}
{Kuehr}, H., {Witzel}, A., {Pauliny-Toth}, I.~I.~K., \& {Nauber}, U. 1981, \bibinfo{title}{{A Catalogue of Extragalactic Radio Sources Having Flux Densities Greater than 1-JY at 5-GHZ},} \aaps, 45, 367

\bibitem[{W.~M. {Lane} {et~al.}(2014){Lane}, {Cotton}, {van Velzen}, {Clarke}, {Kassim}, {Helmboldt}, {Lazio}, \& {Cohen}}]{vlssr}
{Lane}, W.~M., {Cotton}, W.~D., {van Velzen}, S., {et~al.} 2014, \bibinfo{title}{{The Very Large Array Low-frequency Sky Survey Redux (VLSSr)},} \mnras, 440, 327, \dodoi{10.1093/mnras/stu256}

\bibitem[{B.~M. Lasker {et~al.}(2008)Lasker {et~al.}}]{lasker2008}
Lasker, B.~M., {et~al.} 2008, \bibinfo{title}{The Second-Generation Guide Star Catalog: Description and Properties,} AJ, 136, 735

\bibitem[{A.~Y. {Lien} {et~al.}(2025){Lien}, {Krimm}, {Markwardt}, {Oh}, {Marcotulli}, {Mushotzky}, {Collins}, {Barthelmy}, {Baumgartner}, {Cenko}, {Koss}, {Laha}, {Sakamoto}, {Palmer}, \& {Parsotan}}]{bat157}
{Lien}, A.~Y., {Krimm}, H.~A., {Markwardt}, C.~B., {et~al.} 2025, \bibinfo{title}{{The 157 Month Swift/BAT All-sky Hard X-Ray Survey},} \apj, 989, 161, \dodoi{10.3847/1538-4357/ade676}

\bibitem[{T. Liu {et~al.}(2015)Liu {et~al.}}]{liu2015vizierSWXCS}
Liu, T., {et~al.} 2015, \bibinfo{title}{VizieR Online Data Catalog: SWXCS III. Cluster catalog from 2005-2012 Swift data (Liu+, 2015),} VizieR Online Data Catalog

\bibitem[{S.~J. {Maddox} {et~al.}(2018){Maddox}, {Valiante}, {Cigan}, {Dunne}, {Eales}, {Smith}, {Dye}, {Furlanetto}, {Ibar}, {de Zotti}, {Millard}, {Bourne}, {Gomez}, {Ivison}, {Scott}, \& {Valtchanov}}]{hatlasdr2}
{Maddox}, S.~J., {Valiante}, E., {Cigan}, P., {et~al.} 2018, \bibinfo{title}{{The Herschel-ATLAS Data Release 2. Paper II. Catalogs of Far-infrared and Submillimeter Sources in the Fields at the South and North Galactic Poles},} \apjs, 236, 30, \dodoi{10.3847/1538-4365/aab8fc}

\bibitem[{E.~A. {Magnier} {et~al.}(2020){Magnier}, {Schlafly}, {Finkbeiner}, {Tonry}, {Goldman}, {R{\"o}ser}, {Schilbach}, {Casertano}, {Chambers}, {Flewelling}, {Huber}, {Price}, {Sweeney}, {Waters}, {Denneau}, {Draper}, {Hodapp}, {Jedicke}, {Kaiser}, {Kudritzki}, {Metcalfe}, {Stubbs}, \& {Wainscoat}}]{panstarrs-dr2}
{Magnier}, E.~A., {Schlafly}, E.~F., {Finkbeiner}, D.~P., {et~al.} 2020, \bibinfo{title}{{Pan-STARRS Photometric and Astrometric Calibration},} \apjs, 251, 6, \dodoi{10.3847/1538-4365/abb82a}

\bibitem[{A. {Mainzer} {et~al.}(2014){Mainzer}, {Bauer}, {Cutri}, {Grav}, {Masiero}, {Beck}, {Clarkson}, {Conrow}, {Dailey}, {Eisenhardt}, {Fabinsky}, {Fajardo-Acosta}, {Fowler}, {Gelino}, {Grillmair}, {Heinrichsen}, {Kendall}, {Kirkpatrick}, {Liu}, {Masci}, {McCallon}, {Nugent}, {Papin}, {Rice}, {Royer}, {Ryan}, {Sevilla}, {Sonnett}, {Stevenson}, {Thompson}, {Wheelock}, {Wiemer}, {Wittman}, {Wright}, \& {Yan}}]{NEOWISE}
{Mainzer}, A., {Bauer}, J., {Cutri}, R.~M., {et~al.} 2014, \bibinfo{title}{{Initial Performance of the NEOWISE Reactivation Mission},} \apj, 792, 30, \dodoi{10.1088/0004-637X/792/1/30}

\bibitem[{B. Manch {et~al.}(2003)Manch {et~al.}}]{manch2003}
Manch, B., {et~al.} 2003, \bibinfo{title}{SUMSS: The Sydney University Molonglo Sky Survey,} MNRAS, 342, 1117

\bibitem[{R.~N. Manchester {et~al.}(2005)Manchester {et~al.}}]{manchester2005australiaPULSAR}
Manchester, R.~N., {et~al.} 2005, \bibinfo{title}{The Australia telescope national facility pulsar catalogue,} AJ, 129, 1993

\bibitem[{F.~J. {Masci} {et~al.}(2019){Masci}, {Laher}, {Rusholme}, {Shupe}, {Groom}, {Surace}, {Jackson}, {Monkewitz}, {Beck}, {Flynn}, {Terek}, {Landry}, {Hacopians}, {Desai}, {Howell}, {Brooke}, {Imel}, {Wachter}, {Ye}, {Lin}, {Cenko}, {Cunningham}, {Rebbapragada}, {Bue}, {Miller}, {Mahabal}, {Bellm}, {Patterson}, {Juri{\'c}}, {Golkhou}, {Ofek}, {Walters}, {Graham}, {Kasliwal}, {Dekany}, {Kupfer}, {Burdge}, {Cannella}, {Barlow}, {Van Sistine}, {Giomi}, {Fremling}, {Blagorodnova}, {Levitan}, {Riddle}, {Smith}, {Helou}, {Prince}, \& {Kulkarni}}]{ZTF}
{Masci}, F.~J., {Laher}, R.~R., {Rusholme}, B., {et~al.} 2019, \bibinfo{title}{{The Zwicky Transient Facility: Data Processing, Products, and Archive},} \pasp, 131, 018003, \dodoi{10.1088/1538-3873/aae8ac}

\bibitem[{M. {Massardi} {et~al.}(2016){Massardi}, {Bonaldi}, {Bonavera}, {De Zotti}, {Lopez-Caniego}, \& {Galluzzi}}]{paco}
{Massardi}, M., {Bonaldi}, A., {Bonavera}, L., {et~al.} 2016, \bibinfo{title}{{The Planck-ATCA Co-eval Observations project: analysis of radio source properties between 5 and 217 GHz},} \mnras, 455, 3249, \dodoi{10.1093/mnras/stv2561}

\bibitem[{E. Massaro {et~al.}(2015)Massaro {et~al.}}]{massaro20155th5BZCAT}
Massaro, E., {et~al.} 2015, \bibinfo{title}{The 5th edition of the Roma-BZCAT. A short presentation,} Ap\&SS, 357, 1

\bibitem[{D. {McConnell} {et~al.}(2012){McConnell}, {Sadler}, {Murphy}, \& {Ekers}}]{atpmn}
{McConnell}, D., {Sadler}, E.~M., {Murphy}, T., \& {Ekers}, R.~D. 2012, \bibinfo{title}{{ATPMN: accurate positions and flux densities at 5 and 8 GHz for 8385 sources from the PMN survey},} \mnras, 422, 1527, \dodoi{10.1111/j.1365-2966.2012.20726.x}

\bibitem[{N. {Mehrtens} {et~al.}(2012){Mehrtens}, {Romer}, {Hilton}, {Lloyd-Davies}, {Miller}, {Stanford}, {Hosmer}, {Hoyle}, {Collins}, {Liddle}, {Viana}, {Nichol}, {Stott}, {Dubois}, {Kay}, {Sahl{\'e}n}, {Young}, {Short}, {Christodoulou}, {Watson}, {Davidson}, {Harrison}, {Baruah}, {Smith}, {Burke}, {Mayers}, {Deadman}, {Rooney}, {Edmondson}, {West}, {Campbell}, {Edge}, {Mann}, {Sabirli}, {Wake}, {Benoist}, {da Costa}, {Maia}, \& {Ogando}}]{XCS}
{Mehrtens}, N., {Romer}, A.~K., {Hilton}, M., {et~al.} 2012, \bibinfo{title}{{The XMM Cluster Survey: optical analysis methodology and the first data release},} \mnras, 423, 1024, \dodoi{10.1111/j.1365-2966.2012.20931.x}

\bibitem[{A. Merloni {et~al.}(2024)Merloni {et~al.}}]{merloni2024}
Merloni, A., {et~al.} 2024, \bibinfo{title}{The eROSITA All-Sky Survey,} A\&A, 682, A34

\bibitem[{R. {Middei} {et~al.}(2022){Middei}, {Giommi}, {Perri}, {Turriziani}, {Sahakyan}, {Chang}, {Leto}, \& {Verrecchia}}]{nublazar}
{Middei}, R., {Giommi}, P., {Perri}, M., {et~al.} 2022, \bibinfo{title}{{The first hard X-ray spectral catalogue of Blazars observed by NuSTAR},} \mnras, 514, 3179, \dodoi{10.1093/mnras/stac1185}

\bibitem[{M.-A. Miville-Deschênes {et~al.}(2017)Miville-Deschênes {et~al.}}]{miville2017physicalMWMC}
Miville-Deschênes, M.-A., {et~al.} 2017, \bibinfo{title}{Physical properties of molecular clouds for the entire Milky Way disk,} ApJ, 834, 57

\bibitem[{D.~G. {Monet} {et~al.}(2003){Monet}, {Levine}, {Canzian}, {Ables}, {Bird}, {Dahn}, {Guetter}, {Harris}, {Henden}, {Leggett}, {Levison}, {Luginbuhl}, {Martini}, {Monet}, {Munn}, {Pier}, {Rhodes}, {Riepe}, {Sell}, {Stone}, {Vrba}, {Walker}, {Westerhout}, {Brucato}, {Reid}, {Schoening}, {Hartley}, {Read}, \& {Tritton}}]{usno}
{Monet}, D.~G., {Levine}, S.~E., {Canzian}, B., {et~al.} 2003, \bibinfo{title}{{The USNO-B Catalog},} \aj, 125, 984, \dodoi{10.1086/345888}

\bibitem[{P. {Morrissey} {et~al.}(2007){Morrissey}, {Conrow}, {Barlow}, {Small}, {Seibert}, {Wyder}, {Budav{\'a}ri}, {Arnouts}, {Friedman}, {Forster}, {Martin}, {Neff}, {Schiminovich}, {Bianchi}, {Donas}, {Heckman}, {Lee}, {Madore}, {Milliard}, {Rich}, {Szalay}, {Welsh}, \& {Yi}}]{galex}
{Morrissey}, P., {Conrow}, T., {Barlow}, T.~A., {et~al.} 2007, \bibinfo{title}{{The Calibration and Data Products of GALEX},} \apjs, 173, 682, \dodoi{10.1086/520512}

\bibitem[{T. {Mufakharov} {et~al.}(2015){Mufakharov}, {Mingaliev}, {Sotnikova}, {Naiden}, \& {Erkenov}}]{ratan}
{Mufakharov}, T., {Mingaliev}, M., {Sotnikova}, Y., {Naiden}, Y., \& {Erkenov}, A. 2015, \bibinfo{title}{{The observed radio/gamma-ray emission correlation for blazars with the Fermi-LAT and the RATAN-600 data},} \mnras, 450, 2658, \dodoi{10.1093/mnras/stv772}

\bibitem[{T. {Murphy} {et~al.}(2010){Murphy}, {Sadler}, {Ekers}, {Massardi}, {Hancock}, {Mahony}, {Ricci}, {Burke-Spolaor}, {Calabretta}, {Chhetri}, {de Zotti}, {Edwards}, {Ekers}, {Jackson}, {Kesteven}, {Lindley}, {Newton-McGee}, {Phillips}, {Roberts}, {Sault}, {Staveley-Smith}, {Subrahmanyan}, {Walker}, \& {Wilson}}]{at20g}
{Murphy}, T., {Sadler}, E.~M., {Ekers}, R.~D., {et~al.} 2010, \bibinfo{title}{{The Australia Telescope 20 GHz Survey: the source catalogue},} \mnras, 402, 2403, \dodoi{10.1111/j.1365-2966.2009.15961.x}

\bibitem[{A. {Neronov} \& D. {Semikoz}(2025){Neronov} \& {Semikoz}}]{neronov}
{Neronov}, A., \& {Semikoz}, D. 2025, \bibinfo{title}{{Catalog of very-high-energy emitting active galactic nuclei at high Galactic latitudes},} arXiv e-prints, arXiv:2506.08497, \dodoi{10.48550/arXiv.2506.08497}

\bibitem[{R.~P. {Norris} {et~al.}(2021){Norris}, {Marvil}, {Collier}, {Kapi{\'n}ska}, {O'Brien}, {Rudnick}, {Andernach}, {Asorey}, {Brown}, {Br{\"u}ggen}, {Crawford}, {English}, {Rahman}, {Filipovi{\'c}}, {Gordon}, {G{\"u}rkan}, {Hale}, {Hopkins}, {Huynh}, {HyeongHan}, {James Jee}, {Koribalski}, {Lenc}, {Luken}, {Parkinson}, {Prandoni}, {Raja}, {Reiprich}, {Riseley}, {Shabala}, {Sheil}, {Vernstrom}, {Whiting}, {Allison}, {Anderson}, {Ball}, {Bell}, {Bunton}, {Galvin}, {Gupta}, {Hotan}, {Jacka}, {Macgregor}, {Mahony}, {Maio}, {Moss}, {Pandey-Pommier}, \& {Voronkov}}]{emu}
{Norris}, R.~P., {Marvil}, J., {Collier}, J.~D., {et~al.} 2021, \bibinfo{title}{{The Evolutionary Map of the Universe pilot survey},} \pasa, 38, e046, \dodoi{10.1017/pasa.2021.42}

\bibitem[{C.~A. {Onken} {et~al.}(2024){Onken}, {Wolf}, {Bessell}, {Chang}, {Luvaul}, {Tonry}, {White}, \& {Da Costa}}]{skymapper-dr4}
{Onken}, C.~A., {Wolf}, C., {Bessell}, M.~S., {et~al.} 2024, \bibinfo{title}{{SkyMapper Southern Survey: Data release 4},} \pasa, 41, e061, \dodoi{10.1017/pasa.2024.53}

\bibitem[{P. Padovani \& P. Giommi(1995)Padovani \& Giommi}]{padovani1995}
Padovani, P., \& Giommi, P. 1995, \bibinfo{title}{{The Connection between x-ray and radio-selected BL Lacertae objects},} Astrophys. J., 444, 567, \dodoi{10.1086/175631}

\bibitem[{P. {Padovani} {et~al.}(2017){Padovani}, {Alexander}, {Assef}, {De Marco}, {Giommi}, {Hickox}, {Richards}, {Smol{\v{c}}i{\'c}}, {Hatziminaoglou}, {Mainieri}, \& {Salvato}}]{Padovani:2017}
{Padovani}, P., {Alexander}, D.~M., {Assef}, R.~J., {et~al.} 2017, \bibinfo{title}{{Active galactic nuclei: what's in a name?},} \aapr, 25, 2, \dodoi{10.1007/s00159-017-0102-9}

\bibitem[{M.~J. {Page} {et~al.}(2012){Page}, {Brindle}, {Talavera}, {Still}, {Rosen}, {Yershov}, {Ziaeepour}, {Mason}, {Cropper}, {Breeveld}, {Loiseau}, {Mignani}, {Smith}, \& {Murdin}}]{xmmom}
{Page}, M.~J., {Brindle}, C., {Talavera}, A., {et~al.} 2012, \bibinfo{title}{{The XMM-Newton serendipitous ultraviolet source survey catalogue},} \mnras, 426, 903, \dodoi{10.1111/j.1365-2966.2012.21706.x}

\bibitem[{M.~R. Panzera {et~al.}(2003)Panzera {et~al.}}]{panzera2003breraBMW}
Panzera, M.~R., {et~al.} 2003, \bibinfo{title}{The Brera Multi-scale Wavelet ROSAT HRI source catalogue,} A\&A, 399, 351

\bibitem[{R. Piffaretti {et~al.}(2011)Piffaretti {et~al.}}]{piffaretti2011}
Piffaretti, R., {et~al.} 2011, \bibinfo{title}{The MCXC: A meta-catalogue of x-ray detected clusters of galaxies,} A\&A, 534, A109

\bibitem[{ {Planck Collaboration} {et~al.}(2018){Planck Collaboration}, {Akrami}, {Arg{\"u}eso}, {Ashdown}, {Aumont}, {Baccigalupi}, {Ballardini}, {Banday}, {Barreiro}, {Bartolo}, {Basak}, {Benabed}, {Bernard}, {Bersanelli}, {Bielewicz}, {Bonavera}, {Bond}, {Borrill}, {Bouchet}, {Burigana}, {Butler}, {Calabrese}, {Carron}, {Chiang}, {Combet}, {Crill}, {Cuttaia}, {de Bernardis}, {de Rosa}, {de Zotti}, {Delabrouille}, {Delouis}, {Di Valentino}, {Dickinson}, {Diego}, {Ducout}, {Dupac}, {Efstathiou}, {Elsner}, {En{\ss}lin}, {Eriksen}, {Fantaye}, {Finelli}, {Frailis}, {Fraisse}, {Franceschi}, {Frolov}, {Galeotta}, {Galli}, {Ganga}, {G{\'e}nova-Santos}, {Gerbino}, {Ghosh}, {Gonz{\'a}lez-Nuevo}, {G{\'o}rski}, {Gratton}, {Gruppuso}, {Gudmundsson}, {Handley}, {Hansen}, {Herranz}, {Hivon}, {Huang}, {Jaffe}, {Jones}, {Keih{\"a}nen}, {Keskitalo}, {Kiiveri}, {Kim}, {Kisner}, {Krachmalnicoff}, {Kunz}, {Kurki-Suonio}, {L{\"a}hteenm{\"a}ki}, {Lamarre}, {Lasenby}, {Lattanzi}, {Lawrence}, {Levrier}, {Liguori}, {Lilje},
  {Lindholm}, {L{\'o}pez-Caniego}, {Ma}, {Mac{\'\i}as-P{\'e}rez}, {Maggio}, {Maino}, {Mandolesi}, {Mangilli}, {Maris}, {Martin}, {Mart{\'\i}nez-Gonz{\'a}lez}, {Matarrese}, {McEwen}, {Meinhold}, {Melchiorri}, {Mennella}, {Migliaccio}, {Miville-Desch{\^e}nes}, {Molinari}, {Moneti}, {Montier}, {Morgante}, {Natoli}, {Oxborrow}, {Pagano}, {Paoletti}, {Partridge}, {Patanchon}, {Pearson}, {Pettorino}, {Piacentini}, {Polenta}, {Puget}, {Rachen}, {Racine}, {Reinecke}, {Remazeilles}, {Renzi}, {Rocha}, {Roudier}, {Rubi{\~n}o-Mart{\'\i}n}, {Salvati}, {Sandri}, {Savelainen}, {Scott}, {Suur-Uski}, {Tauber}, {Tavagnacco}, {Toffolatti}, {Tomasi}, {Trombetti}, {Tucci}, {Valiviita}, {Van Tent}, {Vielva}, {Villa}, {Vittorio}, {Wehus}, {Zacchei}, \& {Zonca}}]{PCNT}
{Planck Collaboration}, {Akrami}, Y., {Arg{\"u}eso}, F., {et~al.} 2018, \bibinfo{title}{{Planck intermediate results. LIV. The Planck multi-frequency catalogue of non-thermal sources},} \aap, 619, A94, \dodoi{10.1051/0004-6361/201832888}

\bibitem[{G. Principe \&  others.(2018)Principe \& others.}]{principe2018first1FLE}
Principe, G., \& others. 2018, \bibinfo{title}{The first catalog of Fermi-LAT sources below 100 MeV,} A\&A, 618, A22

\bibitem[{H.~P. {Reuter} {et~al.}(1997){Reuter}, {Kramer}, {Sievers}, {Paubert}, {Moreno}, {Greve}, {Leon}, {Panis}, {Ruiz-Moreno}, {Ungerechts}, \& {Wild}}]{mmmonitoring}
{Reuter}, H.~P., {Kramer}, C., {Sievers}, A., {et~al.} 1997, \bibinfo{title}{{Millimetre continuum measurements of extragalactic radio sources. IV. Data from 1993-1994},} \aaps, 122, 271, \dodoi{10.1051/aas:1997333}

\bibitem[{N. {Sahakyan} {et~al.}(2024){Sahakyan}, {Vardanyan}, {Giommi}, {B{\'e}gu{\'e}}, {Israyelyan}, {Harutyunyan}, {Manvelyan}, {Khachatryan}, {Dereli-B{\'e}gu{\'e}}, \& {Gasparyan}}]{MMDC}
{Sahakyan}, N., {Vardanyan}, V., {Giommi}, P., {et~al.} 2024, \bibinfo{title}{{Markarian Multiwavelength Data Center (MMDC): A Tool for Retrieving and Modeling Multitemporal, Multiwavelength, and Multimessenger Data from Blazar Observations},} \aj, 168, 289, \dodoi{10.3847/1538-3881/ad8231}

\bibitem[{D. {Salvetti} {et~al.}(2017){Salvetti}, {Chiaro}, {La Mura}, \& {Thompson}}]{salvetti}
{Salvetti}, D., {Chiaro}, G., {La Mura}, G., \& {Thompson}, D.~J. 2017, \bibinfo{title}{{3FGLzoo: classifying 3FGL unassociated Fermi-LAT {\ensuremath{\gamma}}-ray sources by artificial neural networks},} \mnras, 470, 1291, \dodoi{10.1093/mnras/stx1328}

\bibitem[{R.~D. Saxton {et~al.}(2008)Saxton {et~al.}}]{saxton2008}
Saxton, R.~D., {et~al.} 2008, \bibinfo{title}{The XMM-Newton Slew Survey,} A\&A, 480, 611

\bibitem[{S. {Sazonov} {et~al.}(2024){Sazonov}, {Burenin}, {Filippova}, {Krivonos}, {Arefiev}, {Borisov}, {Buntov}, {Chen}, {Ehlert}, {Garanin}, {Garin}, {Grigorovich}, {Lapshov}, {Levin}, {Lutovinov}, {Mereminskiy}, {Molkov}, {Pavlinsky}, {Ramsey}, {Semena}, {Semena}, {Shtykovsky}, {Sunyaev}, {Tkachenko}, {Swartz}, {Uskov}, {Vikhlinin}, {Voron}, {Zakharov}, \& {Zaznobin}}]{art-xc}
{Sazonov}, S., {Burenin}, R., {Filippova}, E., {et~al.} 2024, \bibinfo{title}{{SRG/ART-XC all-sky X-ray survey: Catalog of sources detected during the first five surveys},} \aap, 687, A183, \dodoi{10.1051/0004-6361/202348950}

\bibitem[{E.~F. Schlafly {et~al.}(2019)Schlafly {et~al.}}]{schlafly2019}
Schlafly, E.~F., {et~al.} 2019, \bibinfo{title}{UnWISE: The unblurred coadds of the WISE imaging,} ApJS, 240, 30

\bibitem[{T.~W. {Shimwell} {et~al.}(2019){Shimwell}, {Tasse}, {Hardcastle}, {Mechev}, {Williams}, {Best}, {R{\"o}ttgering}, {Callingham}, {Dijkema}, {de Gasperin}, {Hoang}, {Hugo}, {Mirmont}, {Oonk}, {Prandoni}, {Rafferty}, {Sabater}, {Smirnov}, {van Weeren}, {White}, {Atemkeng}, {Bester}, {Bonnassieux}, {Br{\"u}ggen}, {Brunetti}, {Chy{\.z}y}, {Cochrane}, {Conway}, {Croston}, {Danezi}, {Duncan}, {Haverkorn}, {Heald}, {Iacobelli}, {Intema}, {Jackson}, {Jamrozy}, {Jarvis}, {Lakhoo}, {Mevius}, {Miley}, {Morabito}, {Morganti}, {Nisbet}, {Orr{\'u}}, {Perkins}, {Pizzo}, {Schrijvers}, {Smith}, {Vermeulen}, {Wise}, {Alegre}, {Bacon}, {van Bemmel}, {Beswick}, {Bonafede}, {Botteon}, {Bourke}, {Brienza}, {Calistro Rivera}, {Cassano}, {Clarke}, {Conselice}, {Dettmar}, {Drabent}, {Dumba}, {Emig}, {En{\ss}lin}, {Ferrari}, {Garrett}, {G{\'e}nova-Santos}, {Goyal}, {G{\"u}rkan}, {Hale}, {Harwood}, {Heesen}, {Hoeft}, {Horellou}, {Jackson}, {Kokotanekov}, {Kondapally}, {Kunert-Bajraszewska}, {Mahatma}, {Mahony}, {Mandal},
  {McKean}, {Merloni}, {Mingo}, {Miskolczi}, {Mooney}, {Nikiel-Wroczy{\'n}ski}, {O'Sullivan}, {Quinn}, {Reich}, {Roskowi{\'n}ski}, {Rowlinson}, {Savini}, {Saxena}, {Schwarz}, {Shulevski}, {Sridhar}, {Stacey}, {Urquhart}, {van der Wiel}, {Varenius}, {Webster}, \& {Wilber}}]{lotss}
{Shimwell}, T.~W., {Tasse}, C., {Hardcastle}, M.~J., {et~al.} 2019, \bibinfo{title}{{The LOFAR Two-metre Sky Survey. II. First data release},} \aap, 622, A1, \dodoi{10.1051/0004-6361/201833559}

\bibitem[{M.~F. {Skrutskie} {et~al.}(2006){Skrutskie}, {Cutri}, {Stiening}, {Weinberg}, {Schneider}, {Carpenter}, {Beichman}, {Capps}, {Chester}, {Elias}, {Huchra}, {Liebert}, {Lonsdale}, {Monet}, {Price}, {Seitzer}, {Jarrett}, {Kirkpatrick}, {Gizis}, {Howard}, {Evans}, {Fowler}, {Fullmer}, {Hurt}, {Light}, {Kopan}, {Marsh}, {McCallon}, {Tam}, {Van Dyk}, \& {Wheelock}}]{2mass}
{Skrutskie}, M.~F., {Cutri}, R.~M., {Stiening}, R., {et~al.} 2006, \bibinfo{title}{{The Two Micron All Sky Survey (2MASS)},} \aj, 131, 1163, \dodoi{10.1086/498708}

\bibitem[{Y. {Stein} {et~al.}(2021){Stein}, {Vollmer}, {Boch}, {Landais}, {Vannier}, {Brouty}, {Allen}, {Derriere}, \& {Ocvirk}}]{specfind}
{Stein}, Y., {Vollmer}, B., {Boch}, T., {et~al.} 2021, \bibinfo{title}{{The SPECFIND V3.0 catalog of radio continuum cross-identifications and spectra: Reaching lower frequencies},} \aap, 655, A17, \dodoi{10.1051/0004-6361/202039659}

\bibitem[{D. {Tripathi} {et~al.}(2024){Tripathi}, {Giommi}, {Di Giovanni}, {Almansoori}, {Al Hamly}, {Arneodo}, {Macci{\`o}}, {Puccetti}, {Barres de Almeida}, {Brandt}, {Di Pippo}, {Doro}, {Israyelyan}, {Pollock}, \& {Sahakyan}}]{firmamento}
{Tripathi}, D., {Giommi}, P., {Di Giovanni}, A., {et~al.} 2024, \bibinfo{title}{{Firmamento: A Multimessenger Astronomy Tool for Citizen and Professional Scientists},} \aj, 167, 116, \dodoi{10.3847/1538-3881/ad216a}

\bibitem[{E. {Valiante} {et~al.}(2016){Valiante}, {Smith}, {Eales}, {Maddox}, {Ibar}, {Hopwood}, {Dunne}, {Cigan}, {Dye}, {Pascale}, {Rigby}, {Bourne}, {Furlanetto}, \& {Ivison}}]{hatlas}
{Valiante}, E., {Smith}, M.~W.~L., {Eales}, S., {et~al.} 2016, \bibinfo{title}{{The Herschel-ATLAS data release 1 - I. Maps, catalogues and number counts},} \mnras, 462, 3146, \dodoi{10.1093/mnras/stw1806}

\bibitem[{W. Voges {et~al.}(2000)Voges {et~al.}}]{voges2000rosatRASS}
Voges, W., {et~al.} 2000, \bibinfo{title}{ROSAT all-sky survey faint source catalogue,} IAU Circ., 7432, 3

\bibitem[{N.~A. Webb {et~al.}(2020)Webb {et~al.}}]{webb2020xmm4XMM}
Webb, N.~A., {et~al.} 2020, \bibinfo{title}{The XMM-Newton serendipitous survey-IX. The fourth XMM-Newton serendipitous source catalogue,} A\&A, 641, A136

\bibitem[{F.~L. Whipple(1966)Whipple}]{whipple1966smithsonianSAO}
Whipple, F.~L. 1966, \bibinfo{title}{Smithsonian astrophysical observatory star catalog,} Smithsonian Astrophysical Observatory Star Catalog

\bibitem[{N.~E. White {et~al.}(2000)White {et~al.}}]{white2000}
White, N.~E., {et~al.} 2000, \bibinfo{title}{The WGACAT version of the ROSAT PSPC catalog,} VizieR Online Data Catalog, IX/31

\bibitem[{R.~L. {White} \& R.~H. {Becker}(1992){White} \& {Becker}}]{north20}
{White}, R.~L., \& {Becker}, R.~H. 1992, \bibinfo{title}{{A New Catalog of 30,239 1.4 GHz Sources},} \apjs, 79, 331, \dodoi{10.1086/191656}

\bibitem[{R.~L. White {et~al.}(1997)White, Becker, Helfand, \& Gregg}]{white1997catalogFIRST}
White, R.~L., Becker, R.~H., Helfand, D.~J., \& Gregg, M.~D. 1997, \bibinfo{title}{A catalog of 1.4 GHz radio sources from the FIRST survey,} ApJ, 475, 479

\bibitem[{A.~E. {Wright} {et~al.}(1994){Wright}, {Griffith}, {Burke}, \& {Ekers}}]{PMN}
{Wright}, A.~E., {Griffith}, M.~R., {Burke}, B.~F., \& {Ekers}, R.~D. 1994, \bibinfo{title}{{The Parkes-MIT-NRAO (PMN) Surveys. II. Source Catalog for the Southern Survey (-87 degrees -4pt.5 < delta < -37 degrees )},} \apjs, 91, 111, \dodoi{10.1086/191939}

\bibitem[{I. {Yamamura} {et~al.}(2010){Yamamura}, {Makiuti}, {Ikeda}, {Fukuda}, {Oyabu}, {Koga}, \& {White}}]{akari}
{Yamamura}, I., {Makiuti}, S., {Ikeda}, N., {et~al.} 2010, \bibinfo{title}{{VizieR Online Data Catalog: AKARI/FIS All-Sky Survey Point Source Catalogues (ISAS/JAXA, 2010)},}, VizieR On-line Data Catalog: II/298. Originally published in: ISAS/JAXA (2010)

\bibitem[{H. {Ye} {et~al.}(2024){Ye}, {Sweijen}, {van Weeren}, {Williams}, {de Jong}, {Morabito}, {Rottgering}, {Shimwell}, {Best}, {Bondi}, {Br{\"u}ggen}, {de Gasperin}, \& {Tasse}}]{lofar}
{Ye}, H., {Sweijen}, F., {van Weeren}, R.~J., {et~al.} 2024, \bibinfo{title}{{1-arcsecond imaging of the ELAIS-N1 field at 144MHz using the LoTSS survey with the international LOFAR telescope},} \aap, 691, A347, \dodoi{10.1051/0004-6361/202348103}

\bibitem[{F. Zwicky {et~al.}(1968)Zwicky, Herzog, \& Wild}]{zwicky1968catalogueZWCLUSTERS}
Zwicky, F., Herzog, E., \& Wild, P. 1968, \bibinfo{title}{Catalogue of Galaxies and of Clusters of Galaxies,} California Institute of Technology (CIT)

\end{thebibliography}
\bibliographystyle{aasjournal}

\appendix
\onecolumngrid

\section{Multi-frequency survey}
\label{app:survey}
The SEDs as well as the multi-wavelength skymaps, in the version of \fshort\ used for this work, are obtained from openly accessible catalogs and survey data, reported in \autoref{tab:catalogs}. Overall we consult 26 catalogs in Radio/Microwave, 9 in Infrared, 10 in Optical/Ultraviolet, 20 in X-rays, 9 in Gamma-rays, and other 16, for a total of 90 catalogs.

\startlongtable
\setlength{\tabcolsep}{3pt} 
\begin{deluxetable}{l l c}
\tablewidth{\columnwidth}
\tabletypesize{\scriptsize}
\tablecaption{\label{tab:catalogs}Catalogs and surveys used by \fshort: Radio--\gray\ and other catalogs. Flag: E = used in \texttt{ERCI}, S = used in generating SED. (Ordered by wavelength range and alphabetical within subgroups).}
\tablehead{
  \colhead{Catalog name and description} & \colhead{Reference} & \colhead{Flag}
}
\startdata
\multicolumn{3}{l}{\textbf{Radio/Microwave}} \\
\tableline
\parbox[t]{8.5cm}{$\square$ ALMA -- ALMA photometry of extragalactic radio sources} & \parbox[t]{6.5cm}{\cite{alma}} & S \\
\parbox[t]{8.5cm}{$\square$ AT20G -- The Australia Telescope 20 GHz Survey} & \parbox[t]{6.5cm}{\cite{at20g}} & S \\
\parbox[t]{8.5cm}{$\square$ ATPMN -- 5 and 8 GHz data from PMN survey} & \parbox[t]{6.5cm}{\cite{atpmn}} & S \\
\parbox[t]{8.5cm}{$\square$ CRATES -- Candidate Radio and AGN Source Survey} & \parbox[t]{6.5cm}{\cite{healey2007cratesCRATES}} & E \\
\parbox[t]{8.5cm}{$\square$ EMU -- Evolutionary Map of the Universe} & \parbox[t]{6.5cm}{\cite{emu}} & E+S \\
\parbox[t]{8.5cm}{$\square$ EPRS -- Extragalactic Radio Sources at Low Frequencies} & \parbox[t]{6.5cm}{\cite{eprs}} & S \\
\parbox[t]{8.5cm}{$\square$ FIRST -- Faint Images of the Radio Sky at Twenty-Centimeters} & \parbox[t]{6.5cm}{\cite{helfand2015lastFIRST, white1997catalogFIRST}} & E+S \\
\parbox[t]{8.5cm}{$\square$ GB6 -- Green Bank 6cm survey} & \parbox[t]{6.5cm}{\cite{GB6}} & E+S \\
\parbox[t]{8.5cm}{$\square$ GB87 -- 87GB Catalog of radio sources} & \parbox[t]{6.5cm}{\cite{GB87}} & S \\
\parbox[t]{8.5cm}{$\square$ GLEAMV2 -- All-sky Murchison Widefield Array (GLEAM) survey} & \parbox[t]{6.5cm}{\cite{gleam}} & S \\
\parbox[t]{8.5cm}{$\square$ KUEHR -- Radio Sources with F\_r $>$ 1Jy} & \parbox[t]{6.5cm}{\cite{kuehr}} & S \\
\parbox[t]{8.5cm}{$\square$ LOFAR -- The LOFAR 120 to 168 MHz observations} & \parbox[t]{6.5cm}{\cite{lofar}} & S \\
\parbox[t]{8.5cm}{$\square$ LoTSS -- The LOFAR Two-metre Sky Survey} & \parbox[t]{6.5cm}{\cite{lotss}} & S \\
\parbox[t]{8.5cm}{$\square$ MM-MONITORING -- mm measurements of extragalactic sources} & \parbox[t]{6.5cm}{\cite{mmmonitoring}} & S \\
\parbox[t]{8.5cm}{$\square$ NORTH20 -- Green Bank 1.4 GHz Northern Sky Survey} & \parbox[t]{6.5cm}{\cite{north20}} & S \\
\parbox[t]{8.5cm}{$\square$ NVSS -- NRAO VLA Sky Survey} & \parbox[t]{6.5cm}{\cite{condon1998nraoNVSS}} & E+S \\
\parbox[t]{8.5cm}{$\square$ PACO -- Planck-ATCA Co-eval Observations} & \parbox[t]{6.5cm}{\cite{paco}} & S \\
\parbox[t]{8.5cm}{$\square$ PCNT -- Planck multi-frequency catalog} & \parbox[t]{6.5cm}{\cite{PCNT}} & S \\
\parbox[t]{8.5cm}{$\square$ PMN -- Parkes-MIT-NRAO survey} & \parbox[t]{6.5cm}{\cite{PMN}} & E+S \\
\parbox[t]{8.5cm}{$\square$ RACS/RACSMID/RACSHIGH -- Rapid ASKAP Continuum low, mid and high frequency surveys} & \parbox[t]{6.5cm}{\cite{hale2021rapidRACS,duchesne2024rapidRACSMID}} & E+S \\
\parbox[t]{8.5cm}{$\square$ RATAN600 -- radio observations of Fermi blazars} & \parbox[t]{6.5cm}{\cite{ratan}} & S \\
\parbox[t]{8.5cm}{$\square$ SPECFIND -- SPECFIND V3 catalog of radio sources} & \parbox[t]{6.5cm}{\cite{specfind}} & S \\
\parbox[t]{8.5cm}{$\square$ SUMSS -- Sydney University Molonglo Sky Survey} & \parbox[t]{6.5cm}{\cite{manch2003}} & E+S \\
\parbox[t]{8.5cm}{$\square$ TEXAS -- Texas Survey of Radio Sources} & \parbox[t]{6.5cm}{\cite{texas}} & S \\
\parbox[t]{8.5cm}{$\square$ VLASSQL -- VLA Sky Survey 3GHz, Quick Look} & \parbox[t]{6.5cm}{\cite{gordon2020catalogVLASSQL}} & E+S \\
\parbox[t]{8.5cm}{$\square$ VLSSR -- VLA Low-frequency Sky Survey Redux} & \parbox[t]{6.5cm}{\cite{vlssr}} & S \\
\parbox[t]{8.5cm}{$\square$ WISH -- Westerbork survey} & \parbox[t]{6.5cm}{\cite{wish}} & S \\
\parbox[t]{8.5cm}{$\square$ TGSS150 -- The GMRT 150 MHz all-sky radio survey} & \parbox[t]{6.5cm}{\cite{tgss}} & S \\
\tableline\tableline
\multicolumn{3}{l}{\textbf{Infrared}} \\
\tableline
\parbox[t]{8.5cm}{$\square$ 2MASS -- Two Micron All Sky Survey} & \parbox[t]{6.5cm}{\cite{2mass}} & E+S \\
\parbox[t]{8.5cm}{$\square$ AKARIBSC -- AKARI/FIS All-Sky Survey Point Source Catalogue} & \parbox[t]{6.5cm}{\cite{akari}} & S \\
\parbox[t]{8.5cm}{$\square$ CatWISE -- WISE} & \parbox[t]{6.5cm}{\cite{CatWISE}} & S \\
\parbox[t]{8.5cm}{$\square$ H-ATLAS-DR1 -- Herschel-ATLAS DR1} & \parbox[t]{6.5cm}{\cite{hatlas}} & S \\
\parbox[t]{8.5cm}{$\square$ H-ATLAS-DR2 -- Herschel-ATLAS DR2 Galactic poles} & \parbox[t]{6.5cm}{\cite{hatlasdr2}} & S \\
\parbox[t]{8.5cm}{$\square$ IRAS-PSC -- IRAS catalogue of Point Sources} & \parbox[t]{6.5cm}{\cite{iras}} & S \\
\parbox[t]{8.5cm}{$\square$ SMARTS -- Opt+IR monitoring of blazars} & \parbox[t]{6.5cm}{\cite{smarts}} & S \\
\parbox[t]{8.5cm}{$\square$ UnWISE -- WISE} & \parbox[t]{6.5cm}{\cite{schlafly2019}} & E+S \\
\parbox[t]{8.5cm}{$\square$ WISE -- Wide-field Infrared Survey Explorer} & \parbox[t]{6.5cm}{\cite{cutri2012vizierWISE-VIZIER}} & S \\
\tableline\tableline
\multicolumn{3}{l}{\textbf{Optical/Ultraviolet}} \\
\tableline
\parbox[t]{8.5cm}{$\square$ 6DF -- Six-degree Field Galaxy Survey} & \parbox[t]{6.5cm}{\cite{jones2009}} & S \\
\parbox[t]{8.5cm}{$\square$ GAIA3 -- Gaia DR3} & \parbox[t]{6.5cm}{\cite{gaiadr3}} & E+S \\
\parbox[t]{8.5cm}{$\square$ GALEX -- Galaxy Evolution Explorer UV catalog} & \parbox[t]{6.5cm}{\cite{galex}} & S \\
\parbox[t]{8.5cm}{$\square$ HSTGSC -- Hubble Guide Star Catalog} & \parbox[t]{6.5cm}{\cite{lasker2008}} & E+S \\
\parbox[t]{8.5cm}{$\square$ PanSTARRS -- PanSTARRS DR2} & \parbox[t]{6.5cm}{\cite{panstarrs-dr2}} & E+S \\
\parbox[t]{8.5cm}{$\square$ SDSS -- Sloan Digital Sky Survey DR19} & \parbox[t]{6.5cm}{\cite{blanton2017sloanSDSS, abolfathi2018fourteenthSDSS}} & E+S \\
\parbox[t]{8.5cm}{$\square$ Skymapper -- Skymapper Southern Survey Data Release 4} & \parbox[t]{6.5cm}{\cite{skymapper-dr4}} & S \\
\parbox[t]{8.5cm}{$\square$ SWIFTUVOT-MMDC -- Swift optical/UV monitor} & \parbox[t]{6.5cm}{\cite{MMDC}} & S \\
\parbox[t]{8.5cm}{$\square$ USNO -- The USNO-B Catalog} & \parbox[t]{6.5cm}{\cite{usno}} & E+S \\
\parbox[t]{8.5cm}{$\square$ XMMOM -- XMM optical monitor} & \parbox[t]{6.5cm}{\cite{xmmom}} & S \\
\tableline\tableline
\multicolumn{3}{l}{\textbf{X-rays}} \\
\tableline
\parbox[t]{8.5cm}{$\square$ 2SXPS -- Swift-XRT Point Source Catalog} & \parbox[t]{6.5cm}{\cite{evans2020}} & E+S \\
\parbox[t]{8.5cm}{$\square$ 4XMM-DR14 -- XMM-Newton 4th Source Catalog} & \parbox[t]{6.5cm}{\cite{webb2020xmm4XMM}} & E+S \\
\parbox[t]{8.5cm}{$\square$ 1OUSX -- 1st Open Universe Soft X-ray Catalog} & \parbox[t]{6.5cm}{\cite{giommi_prep}} & E+S \\
\parbox[t]{8.5cm}{$\square$ BAT157 -- Swift-BAT 157 months} & \parbox[t]{6.5cm}{\cite{bat157}} & S \\
\parbox[t]{8.5cm}{$\square$ BeppoSAX -- BeppoSAX spectra of blazars} & \parbox[t]{6.5cm}{\cite{bepposax}} & S \\
\parbox[t]{8.5cm}{$\square$ BMW -- Brera Multi-scale Wavelet ROSAT HRI Catalog} & \parbox[t]{6.5cm}{\cite{panzera2003breraBMW}} & E+S \\
\parbox[t]{8.5cm}{$\square$ CSC2.1 -- Chandra Source Catalog Version 2.1} & \parbox[t]{6.5cm}{\cite{evans2010}} & E+S \\
\parbox[t]{8.5cm}{$\square$ eFEDS -- eROSITA Final Equatorial Depth Survey} & \parbox[t]{6.5cm}{\cite{brunner2022erositaEFEDS}} & E+S \\
\parbox[t]{8.5cm}{$\square$ eRASS1 -- eROSITA All-Sky Survey} & \parbox[t]{6.5cm}{\cite{merloni2024}} & E+S \\
\parbox[t]{8.5cm}{$\square$ eRASS1-S -- eROSITA All-Sky Survey South Subset} & \parbox[t]{6.5cm}{\cite{merloni2024}} & E+S \\
\parbox[t]{8.5cm}{$\square$ IPC2E -- Einstein IPC X-ray Source Catalog} & \parbox[t]{6.5cm}{\cite{harris1990einsteinIPC2E}} & E+S \\
\parbox[t]{8.5cm}{$\square$ IPCSL -- Einstein IPC Slew Survey} & \parbox[t]{6.5cm}{\cite{elvis1992}} & S \\
\parbox[t]{8.5cm}{$\square$ NuBlazar -- Open Universe NuSTAR blazars spectra} & \parbox[t]{6.5cm}{\cite{nublazar}} & S \\
\parbox[t]{8.5cm}{$\square$ RASS -- ROSAT All-Sky Survey} & \parbox[t]{6.5cm}{\cite{voges2000rosatRASS}} & E+S \\
\parbox[t]{8.5cm}{$\square$ SRG/ART-XC all-sky X-ray survey} & \parbox[t]{6.5cm}{\cite{art-xc}} & S \\
\parbox[t]{8.5cm}{$\square$ SUFST -- Swift-XRT ultra-fast analysis} & \parbox[t]{6.5cm}{Casotto et al.\ 2025} & S \\
\parbox[t]{8.5cm}{$\square$ SWIFTXRT-MMDC -- Swift XRT spectra of blazars} & \parbox[t]{6.5cm}{\cite{MMDC}} & S \\
\parbox[t]{8.5cm}{$\square$ SWXCS -- Swift X-ray Cluster Survey} & \parbox[t]{6.5cm}{\cite{liu2015vizierSWXCS}} & E \\
\parbox[t]{8.5cm}{$\square$ XCS -- XMM X-Ray Cluster Survey} & \parbox[t]{6.5cm}{\cite{XCS}} & E \\
\parbox[t]{8.5cm}{$\square$ XMMSL3 -- XMM-Newton Slew Survey Catalog DR3} & \parbox[t]{6.5cm}{\cite{saxton2008}} & E+S \\
\parbox[t]{8.5cm}{$\square$ WGACAT -- ROSAT PSPC Catalog} & \parbox[t]{6.5cm}{\cite{white2000}} & E+S \\
\tableline\tableline
\multicolumn{3}{l}{\textbf{Gamma-rays}} \\
\tableline
\parbox[t]{8.5cm}{$\square$ 1FLE -- \lat\ catalog below 100 MeV} & \parbox[t]{6.5cm}{\cite{principe2018first1FLE}} & S \\
\parbox[t]{8.5cm}{$\square$ 2AGILE -- AGILE 2nd Gamma-ray Source Catalog} & \parbox[t]{6.5cm}{\cite{bulgarelli2019vizier2AGILE}} & S \\
\parbox[t]{8.5cm}{$\square$ 2BIGB -- Catalog of HSP \gray\, blazars} & \parbox[t]{6.5cm}{\cite{2bigb}} & S \\
\parbox[t]{8.5cm}{$\square$ 3FHL -- \lat\ 3rd Hard Source Catalog} & \parbox[t]{6.5cm}{\cite{ajello20173fhl3FHL}} & E+S \\
\parbox[t]{8.5cm}{$\square$ 4FGL-DR3 -- \lat\ 4th Source Catalog DR3} & \parbox[t]{6.5cm}{\cite{4FGL-DR3}} & S \\
\parbox[t]{8.5cm}{$\square$ 4FGL-DR4 -- \lat\ 4th Source Catalog DR4} & \parbox[t]{6.5cm}{\cite{4FGL-DR4}} & E+S \\
\parbox[t]{8.5cm}{$\square$ 4LAC-DR3 -- \lat\ 4th AGN Catalog} & \parbox[t]{6.5cm}{\cite{4LAC-DR3}} & E \\
\parbox[t]{8.5cm}{$\square$ FERMI-LAT-MMDC -- Fermi-LAT spectra} & \parbox[t]{6.5cm}{\cite{MMDC}} & S \\
\parbox[t]{8.5cm}{$\square$ MAGIC -- Spectral data from selected papers} & \parbox[t]{6.5cm}{\cite{Doro:2019zvt}} & S \\
\tableline\tableline
\multicolumn{3}{l}{\textbf{Multiwavelength / Known sources / Other}} \\
\tableline
\parbox[t]{8.5cm}{$\square$ 3HSP -- High Synchrotron Peaked Blazar Catalog} & \parbox[t]{6.5cm}{\cite{chang20193hsp3HSP}} & E \\
\parbox[t]{8.5cm}{$\square$ 5BZCat -- 5th edition of the Roma-BZCAT} & \parbox[t]{6.5cm}{\cite{massaro20155th5BZCAT}} & E \\
\parbox[t]{8.5cm}{$\square$ ABELL -- Abell Catalog of Rich Clusters of Galaxies} & \parbox[t]{6.5cm}{\cite{abell1989}} & E \\
\parbox[t]{8.5cm}{$\square$ CVCAT -- Cataclysmic Variable Catalog} & \parbox[t]{6.5cm}{\cite{kube2003cvcatCVCAT}} & E \\
\parbox[t]{8.5cm}{$\square$ Fermi3PSR -- Third \lat\ Gamma-Ray Pulsar Catalog} & \parbox[t]{6.5cm}{\cite{abdo2013secondF2PSR}} & E \\
\parbox[t]{8.5cm}{$\square$ MCXC -- Catalog of X-ray detected Clusters} & \parbox[t]{6.5cm}{\cite{piffaretti2011}} & E \\
\parbox[t]{8.5cm}{$\square$ PULSAR -- ATNF Pulsar Catalog} & \parbox[t]{6.5cm}{\cite{manchester2005australiaPULSAR}} & E \\
\parbox[t]{8.5cm}{$\square$ PSZ2 -- Second Planck SZ Catalog} & \parbox[t]{6.5cm}{\cite{planck2016}} & E \\
\parbox[t]{8.5cm}{$\square$ RASS -- ROSAT All-Sky Survey} & \parbox[t]{6.5cm}{\cite{voges2000rosatRASS}} & E \\
\parbox[t]{8.5cm}{$\square$ SAO -- Smithsonian Astrophysical Observatory Star Catalog} & \parbox[t]{6.5cm}{\cite{whipple1966smithsonianSAO}} & E \\
\parbox[t]{8.5cm}{$\square$ SNRGREEN -- Green's Supernova Remnant Catalog} & \parbox[t]{6.5cm}{\cite{green2025updatedSNRGREEN}} & E \\
\parbox[t]{8.5cm}{$\square$ SPTSZ -- SPT-SZ Survey} & \parbox[t]{6.5cm}{\cite{sptsz}} & E \\
\parbox[t]{8.5cm}{$\square$ MilliQuas -- Million Quasars Catalog} & \parbox[t]{6.5cm}{\cite{flesch2015}} & E \\
\parbox[t]{8.5cm}{$\square$ MWMC -- Milky Way Molecular Clouds} & \parbox[t]{6.5cm}{\cite{miville2017physicalMWMC}} & E \\
\parbox[t]{8.5cm}{$\square$ MWSC -- Milky Way Stellar Clusters} & \parbox[t]{6.5cm}{\cite{kharchenko2013globalMWSC}} & E \\
\parbox[t]{8.5cm}{$\square$ XRBCAT -- X-ray binaries catalog} & \parbox[t]{6.5cm}{\cite{avakyan2023xrbcatsXRBCAT}} & E \\
\parbox[t]{8.5cm}{$\square$ ZWCLUSTERS -- Zwicky Cluster Catalog} & \parbox[t]{6.5cm}{\cite{zwicky1968catalogueZWCLUSTERS}} & E \\
\tableline\tableline
\enddata
\end{deluxetable}

\onecolumngrid
\section{Step by step procedure for source association}
\label{app:step_by_step}
Hereafter we list the step-by-step procedure used for source association for this work

\begin{enumerate}
    \item From the \fshort\ portal, load the 4FGL catalog of high-latitude sources with \texttt{Data Access--> User Input --> Import a new table} command, see \autoref{fig:AR_step_1}. The file, properly formatted contain information such as source name, positional error, association name.
    \begin{figure}[h!t]
    \centering
    \includegraphics[width=0.6\columnwidth]{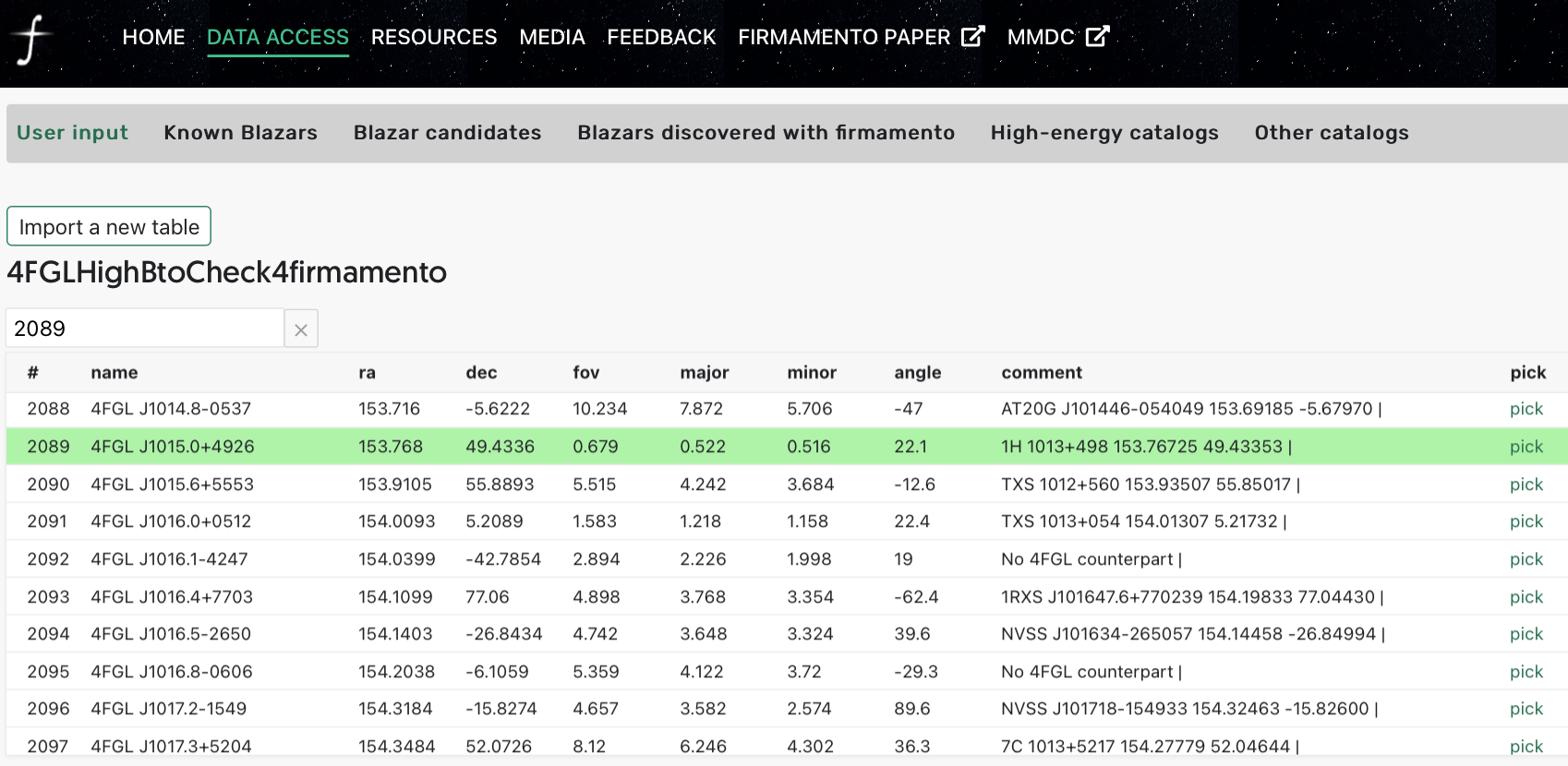}
    \caption{Table for 5,063 \lat\ 4FGL catalog sources of interest, uploaded on the \fshort\ platform for counterpart search.}
    \label{fig:AR_step_1} 
    \end{figure}
\item By using the command \texttt{pick}, select the source. This automatically launches the ERCI algorithm for candidate identification. Because \fshort\ stores previous searches on a specific source direction, it is possible to \texttt{Force run} to re-run the procedure from scratch. During this step, freely accessible multi-wavelength online catalogs are parsed. In case the catalog is not available during a first call, further two calls are made on the specific catalog. The availability of external catalogs is not always guaranteed by their providers; for this reason, \fshort\ is evolving toward a version in which the multi-wavelength catalogs are, whenever possible, stored locally.
As an output, \fshort\ provides zero, one or more plausible candidate associations. This is shown in \autoref{fig:step_by_step1} (left).

        \begin{figure}[h!t]
    \centering
\includegraphics[width=0.32\columnwidth]{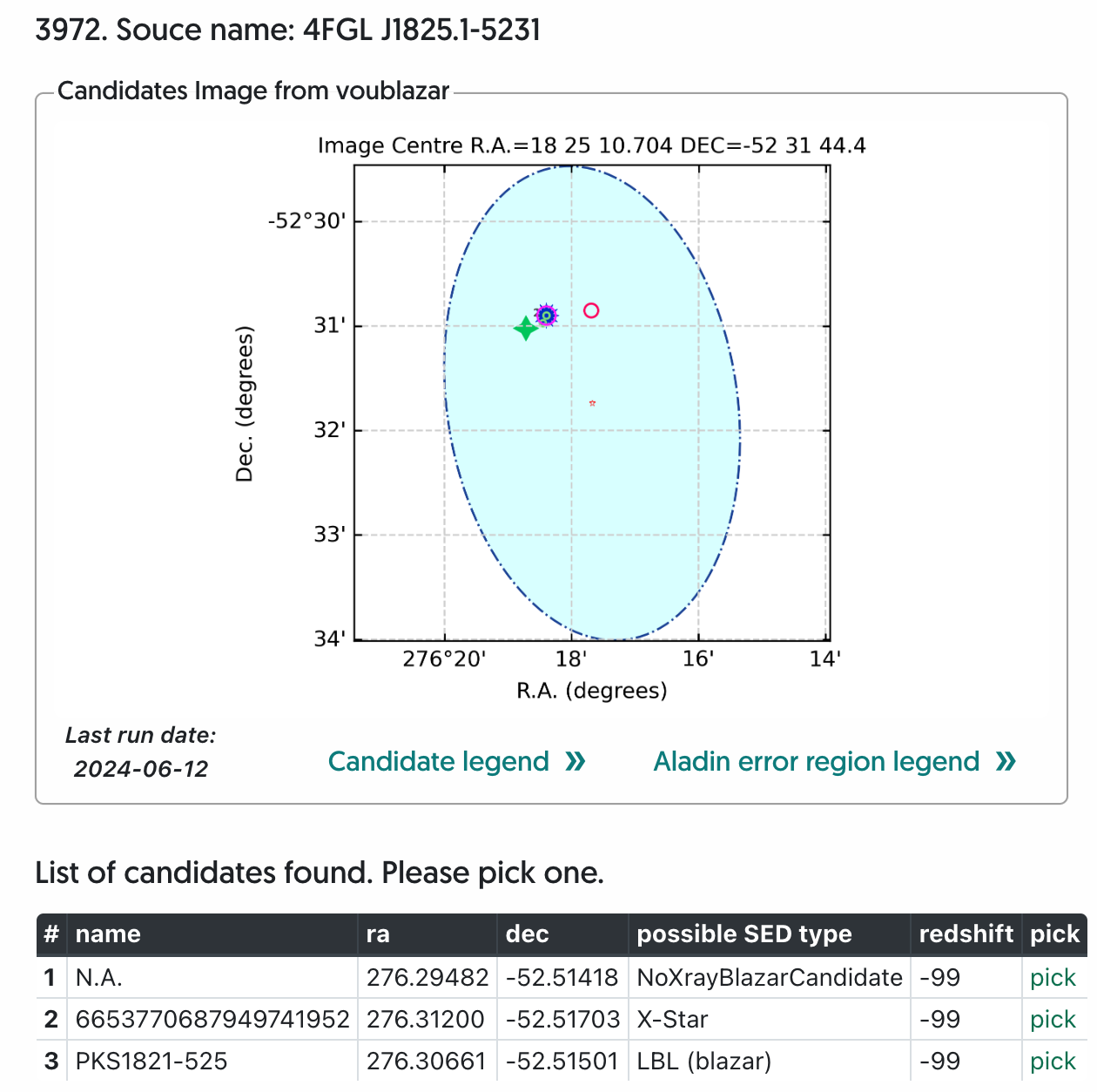} \hfill
\includegraphics[width=0.32\columnwidth]{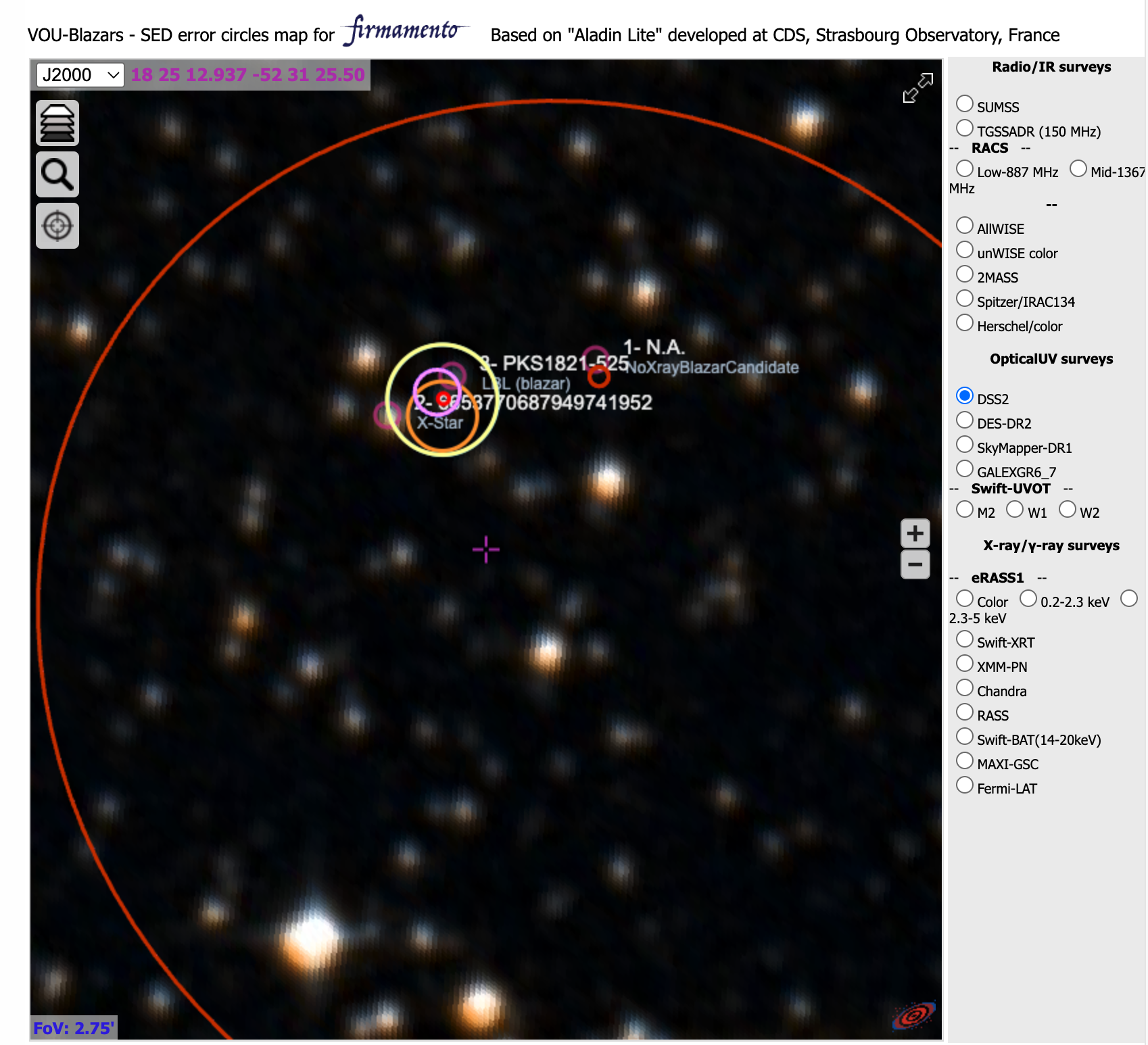} \hfill
\includegraphics[width=0.32\columnwidth]{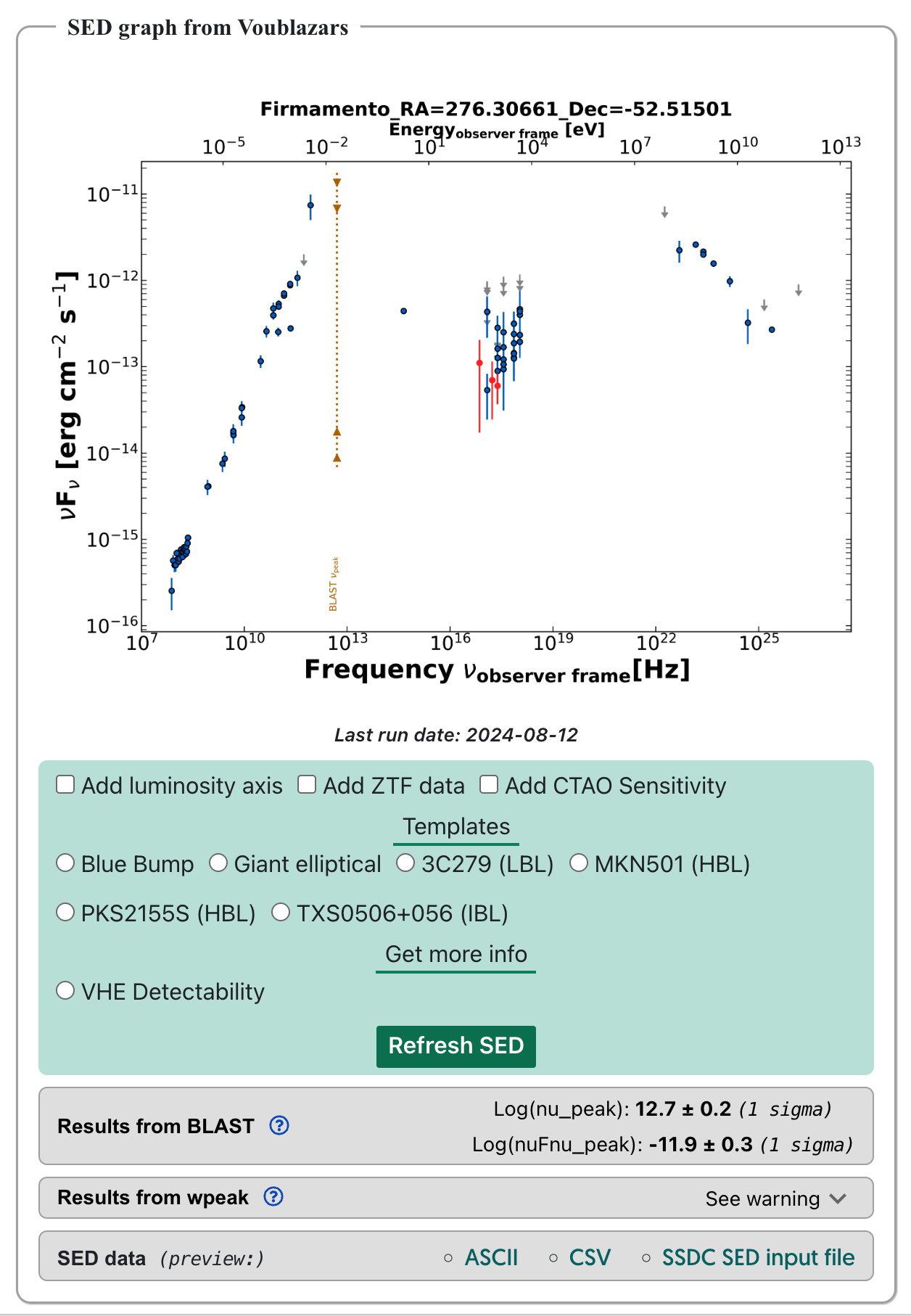} 
    \caption{Example of \fshort\ graphical view after a tentative association 4FGL J1825.1-5231. (left) The candidate 4FGL J1825.1-5231 is picked from a list (ID 3972), ERCI returns three candidates out of which only one is putative blazar, the 3rd one, PKS1821-525. (center) An aladin skymap shows the multi-wavelength morphology with all available catalog error circles. The yellow circle is a 4LAC source. Candidate 3 therefore agrees with 4LAC. (right) Multi-wavelength SED generated by \fshort\ on the selected association}
    \label{fig:step_by_step1} 
    \end{figure}
        \item A comparison is made with LAT catalogs (4FGL, 4LAC) and sources are classified based on the LAT proposed localization, the \fshort\ proposed location, the check of SEDs at these localization (using option \texttt{Get SED data}) and if needed by the inspection of the \texttt{Aladin} multi-wavelength skymaps. See \autoref{fig:step_by_step1} (center).
    \item On the select source, \texttt{Get SED data} automatically run \texttt{blast} and \texttt{wpeak}. The output values of the synchrotron peak and flux are recorded. See \autoref{fig:step_by_step1} (right).
\end{enumerate}

\onecolumngrid
\section{4LAC-DR3 sources not in 4FGL-DR4}
\label{app:4lac_excluded}
The list of sources in 4LAC not present in 4FGL-DR4 includes 24 sources:
\begin{itemize}
    \item 12 sources have different naming in 4FGFL-DR3 than 4FGL-DR4 probably due to updated sky coordinates. For example, 4FGL J0301.6-7155, listed as such in 4FGL-DR3, is listed as 4FGL J0301.5-7156 in 4FGL-DR4. Their complete list is (4FGL label omitted): \texttt{CLASS1="fsrq"}: J0301.6-7155, J1423.5-7829; J2207.5-5346; \texttt{CLASS1="rdg"}:
J0322.6-3712e, J1324.0-4330e; \texttt{CLASS1="bcu"}: 
J0430.2-0356, J0623.7-3348, J0728.0+6735, J1416.1+1320; \texttt{CLASS1="bll"}: J2346.7+0705, J2236.6+3706; J2317.4+4533
\item 12 sources appear in 4FGL-DR3 with "c" (confusion) letter after the name (4FGL label omitted): J0344.2+3203c, J0506.0-0357c, J0517.9-6930c, J0521.8+5658c, J0535.7-6604c, J0539.7-0521c, J0545.0+0613c, J0554.3-1009c, J0733.7+0205c, J0743.3-4912c, J1644.8-2154c, J2108.7+7532c. They have all \texttt{CLASS1="bcu"}.
\end{itemize}

Because we could not find an automatic procedure, and their number is small compared to the 4FGL-DR3/4LAC blazars, we neglected their screening with \fshort\ at this time.

\begin{deluxetable}{lcll}
\tabletypesize{\footnotesize}
\tablecaption{Description of the \texttt{1FLAT.fits} binary table columns. Catalog options in \autoref{tab:tags}\label{tab:1FLAT_fits_description}}
\tablehead{
\colhead{Column} & \colhead{Format} & \colhead{Unit} & \colhead{Description}
}
\startdata
\cutinhead{\fcat\, \fshort\ related entries $\downarrow$}
1FLAT\_name         & 24A   &                & 1FLAT JHHMMSS.f+/-DDMMSS source name \\
                        &    &              & 4GFL source name  for unassociated \\
RAJ2000               & E     & deg            & Right ascension (J2000) of \fshort\ association \\
                      &      &             & Right ascension (J2000) of 4GFL-DR4 for unassociated \\
DEJ2000               & E     & deg            & Declination (J2000) of \fshort\ association \\
                      &      &             & Declination (J2000) of 4GFL-DR4 for unassociated \\
CLASS                 & 6A    &                & Class designation for associated source \\
nu\_syn               & E     & Hz             & Synchrotron-peak frequency from \texttt{wpeak} (observer frame; $\log_{10}$) \\
nuFnu\_syn            & E     & erg\,cm$^{-2}$\,s$^{-1}$ & $\nu F\nu$ at synchrotron peak from \texttt{wpeak} (observer frame; $\log_{10}$) \\
TAG                   & 30A   &                & Classification tag \\
\cutinhead{\lat\ 4FGL-DR4 related entries $\downarrow$}
4FGL\_Source\_Name          & 18A   &                & Source name 4FGL JHHMM.f+DDMM (4FGL designation) \\
4FGL\_ASSOC1          & 30A   &                & Name of identified or likely associated source (primary) \\
4FGL\_CLASS1          & 6A    &                & Class designation for associated source (primary) \\
4FGL\_ASSOC2          & 30A   &                & Name of identified or likely associated source (secondary) \\
4FGL\_CLASS2          & 6A    &                & Class designation for associated source (secondary) \\
4FGL\_Signif\_Avg     & E     &                & Source significance in $\sigma$ over 50\,MeV–1\,TeV \\
4FGL\_PL\_Index       & E     &                & Photon index from PowerLaw fit \\
4FGL\_Energy\_Flux100 & E     & erg\,cm$^{-2}$\,s$^{-1}$ & Energy flux 100\,MeV–100\,GeV from spectral fit \\
4FGL\_Frac\_Variability & E   &                & Fractional variability index \\
\cutinhead{\lat\ 4LAC-DR3 related entries $\downarrow$}
4LAC\_ASSOC1          & 30A   &                & Name of identified or likely associated source \\
4LAC\_CLASS           & 6A    &                & Class designation for associated source \\
4LAC\_nu\_syn         & E     & Hz             & Synchrotron-peak frequency (observer frame; $\log_{10}$) \\
4LAC\_nuFnu\_syn      & E     & erg\,cm$^{-2}$\,s$^{-1}$ & $\nu F\nu$ at synchrotron peak (observer frame; $\log_{10}$) \\
\enddata
\end{deluxetable}

\onecolumngrid
\section{Description of the FITS Version of the \fcat\, Catalog.}
\label{sec:1FLAT_fits}

The catalog has been released in FITS format, following standard conventions for
binary table extensions. The file \texttt{\fcat.fits} is produced from the original
CSV table and includes all relevant fields describing the source properties. During the conversion, column names have been sanitized to ensure compatibility with the FITS standard (only uppercase/lowercase letters, digits, and underscores
are used). The final table therefore provides a clean dataset suitable
for scientific analysis.

The FITS file is structured as a binary table in the first extension (HDU~1).
Each row corresponds to a source entry, and each column contains a catalog
parameter such as identifiers, associations, spectral properties, variability,
and classification tags. Metadata such as column names and types are stored in
the FITS header. The data can be easily accessed using common astronomical
software libraries such as \texttt{Astropy} in Python, or visualized with FITS
viewers. A detailed description of the individual columns, together with their units, is
provided in \autoref{tab:1FLAT_fits_description}. In \autoref{tab:tags} we report the labels taken by the entries \texttt{TAG} and \texttt{CLASS}.

\noindent

\begin{deluxetable*}{ll}
\tablecaption{Values for classification \texttt{TAG, CLASS} used in \fcat\label{tab:tags} }
\tablehead{
\colhead{Column/Options} & \colhead{Description}
}
\startdata
\hline
\texttt{TAG} & \\
\hline
\texttt{BLAZAR confirmed}         & For blazars in which \fshort\ agrees with \lat\ in unassisted way \\
\texttt{BLAZAR confirmed visual} & For blazars in which \fshort\ agrees with \lat\ in assisted way \\
\texttt{BLAZAR different}      & For blazars in which \fshort\ provides an alternative association to 4FGL or 4LAC \\
\texttt{BLAZAR new}       & Blazars discovered with \fshort\ and not found in 4FGL or 4LAC \\
\texttt{UNCERTAIN}       & For sources in which \fshort\ does not find a valid candidate whereas \lat\ does \\
\texttt{NOC confirmed}            & For sources in which \fshort\ agrees with \lat\ that no counterparts are found \\
\texttt{NOC new}             & For sources where \fshort\ does not find a counterpart whereas \lat\ does \\
\texttt{GALAXY confirmed}    & For galaxies in which \fshort\ agrees with \lat \\
\texttt{GALAXY new}          & For galaxies in which \fshort\ disagrees with \lat \\
\hline\hline
\texttt{CLASS} & \\
\hline
\texttt{HSP}        & $\log_{10}(\nu_{\text{peak}}/\mathrm{Hz}) \geq 15$ \\
\texttt{IBL}        & $13.5 \leq \log_{10}(\nu_{\text{peak}}/\mathrm{Hz}) < 15$ \\
\texttt{LSP}  & $\log_{10}(\nu_{\text{peak}}/\mathrm{Hz}) < 13.5$ \\
\texttt{noC}        & No candidate found by \fshort \\
\texttt{galaxy}  & A galaxy \\
\texttt{uncertain}  & Uncertain classification \\
\enddata
\end{deluxetable*}
\end{document}